\documentclass[runningheads]{llncs}
\usepackage[T1]{fontenc}
\usepackage{graphicx}
\usepackage{float}
\usepackage{multirow}
\usepackage{tabularray}
\usepackage{tikz}
\usetikzlibrary{patterns}
\usepackage{float}
\usepackage{xspace}
\usepackage{xcolor}
\usepackage{colortbl}
\usepackage{booktabs}
\usepackage{threeparttable}
\usepackage{bm}
\usepackage{amsfonts,amssymb,amsbsy,amsmath}
\usepackage{graphicx}
\usepackage{algorithm}
\usepackage{algpseudocode}
\usepackage[colorlinks=true, linkcolor=blue, citecolor=blue, urlcolor=blue]{hyperref}
\usepackage{bbding}
\usepackage{academicons}
\newcommand{\orcid}[1]{\href{https://orcid.org/#1}{\textcolor[HTML]{A6CE39}{\aiOrcid}}}

\usepackage{bold-extra}
\newcommand\skinny{\texttt{SKINNY}\xspace}
\newcommand{\skinnyver}[2]{\texttt{SKINNY-#1-#2}}
\newcommand\speck{\textsc{Speck}}

\newcommand\etal{\textit{et al.}}

\newtheorem{rrule}{Reducing Rule}
\newtheorem{pattern}{Pattern}
\newtheorem{distinguisher}{Distinguisher}

\newcommand{\fillcolor}[1]{\relax \tikz{\draw[solid, fill = {#1}, scale = 0.18] (0, 0) rectangle (1, 1);}}

\newcommand{\Xcell}{\relax \tikz{\draw[solid, pattern = north east lines, scale = 0.2] (0, 0) rectangle (1, 1);}\xspace}

\begin{document}

\title{Neural-Inspired Advances in Integral Cryptanalysis}
\titlerunning{Neural-Inspired Advances in Integral Cryptanalysis}

\author{Liu Zhang\inst{2,3} \and
Yiran Yao\inst{3} \and
Danping Shi\inst{1}\textsuperscript{\Envelope} \and Dongchen Chai\inst{2} \and
Jian Guo\inst{3} \and
Zilong Wang\inst{2} \textsuperscript{\Envelope}
        }
\institute{
        Key Laboratory of Cyberspace Security Defense,
        Institute of Information Engineering, Chinese Academy of Sciences \email{shidanping@iie.ac.cn}\and
        School of Cyber Engineering, Xidian University, Xi'an, China \email{liu.zhang@ntu.edu.sg,chaidc@foxmail.com,zlwang@xidian.edu.cn} \and
	   School of Physical and Mathematical Sciences, Nanyang Technological University, Singapore
        \email{\{yiran005@e,guojian@\}.ntu.edu.sg}
}

\maketitle              

\begin{abstract}
The study by Gohr \etal{} at  CRYPTO~2019 and sunsequent related works have shown that neural networks can uncover previously unused features, offering novel insights into cryptanalysis. Motivated by these findings, we employ neural networks to learn features specifically related to integral properties and integrate the corresponding insights into optimized search frameworks. These findings validate the framework of using neural networks for feature exploration, providing researchers with novel insights that advance established cryptanalysis methods.
\par
Neural networks have inspired the development of more precise integral search models. By comparing the integral distinguishers obtained via neural networks with those identified by classical methods, we observe that existing automated search models often fail to find optimal distinguishers. To address this issue, we develop a meet-in-the-middle search framework that balances model accuracy and computational efficiency. As a result, we reduce the number of active plaintext bits required for an 11-round integral distinguisher on \texttt{SKINNY-64-64}, and further identify a 12-round key-dependent integral distinguisher—achieving one additional round over the previous best-known result.
\par\par
The integral distinguishers discovered by neural networks enable key-recovery attacks on more rounds. We identify a 7-round key-independent integral distinguisher from neural networks with even only one active plaintext cell, which is based on linear combinations of bits. This distinguisher enables a 15-round key-recovery attack on \skinnyver{n}{n} through a strategy with 3 rounds of forward decryption and 5 rounds of backward encryption, improving upon the previous record by one round. The same distinguisher also enhances attacks on \skinnyver{n}{2n} and \skinnyver{n}{3n}. Additionally, we discover an 8-round key-dependent integral distinguisher using neural network that further reduces the time complexity of key-recovery attacks against \texttt{SKINNY}.

\end{abstract}

\section{Introduction}
Integral cryptanalysis, first introduced by Knudsen in~\cite{conf/fse/KnudsenW02}, analyzes the algebraic structure of a block cipher by computing the sum of ciphertexts derived from a set of plaintexts that span a linear subspace. It tracks how integral properties—\texttt{ALL}, \texttt{BALANCE}, \texttt{CONSTANT}, and \texttt{UNKNOWN}—propagate through the internal state words via the cipher’s round transformations.
The division property, originally proposed in~\cite{conf/eurocrypt/Todo15}, provides a more general and precise framework for identifying integral distinguishers. For a set of texts $\mathbb{X} \subseteq \mathbb{F}_2^n$, the division property is described using a subset of indicator vectors $u \in \mathbb{F}_2^n$, which is divided into two categories: the 0-subset and the $unknown$-subset.
To further refine the granularity of integral analysis, the three-subset bit-based division property was introduced in~\cite{conf/fse/TodoM16}. In this approach, the set of $u$ is partitioned into three subsets: the 0-subset, the $unknown$-subset, and the 1-subset.
Also in 2016, Boura and Canteaut introduced the notion of parity sets~\cite{crypto/BouraC16}, offering an alternative view of the division property by connecting it with the algebraic normal form (ANF) of the cipher's components.
Lambin \etal{} extended the analysis by considering ciphers with linearly equivalent S-boxes and successfully identified high-round integral distinguishers for the \texttt{RECTANGLE}~\cite{dcc/LambinDF20}.
More recently, in FSE 2023, Beyne and Verbauwhede  introduced the notion of generalized integral properties, which not only consider mapping-based integral properties on ciphertext sets, but also take into account transformations applied to the plaintext inputs~\cite{tosc/BeyneV23}. The above discussions focus on deterministic integral properties. In contrast, probabilistic integral properties, which correspond to the cube tester, were introduced in~\cite{fse/AumassonDMS09}.

\par
One primary challenge in discovering integral distinguishers lies in effectively modeling the propagation of the division property through the round functions of a cipher. The propagation was initially evaluated using a breadth-first search algorithm in~\cite{conf/eurocrypt/Todo15,conf/fse/TodoM16,joc/Todo17}, but this approach becomes computationally infeasible for block ciphers with large state sizes.
To improve efficiency, Xiang \etal{} introduced the concept of division trails and proposed a MILP-based automatic search method~\cite{asiacrypt/xiang2016applying}. Sun \etal{} developed an alternative framework based on Boolean Satisfiability Problem (SAT) to analyze ARX ciphers~\cite{asiacrypt/SunWW17}.
Wang \etal{} further refined the MILP-based approach by accurately modeling the three-subset division property and incorporating pruning techniques and fast propagation methods~\cite{conf/asiacrypt/WangHGZS19}. Hu \etal{} introduced the monomial prediction technique, which accelerates the enumeration of monomial trials and enables the recovery of more superpolies~\cite{asiacrypt/HuSW020}.
Lambin \etal{} extended ciphers with linear mappings and successfully identified high-round integral distinguishers based on linear combinations of bits for \texttt{RECTANGLE}~\cite{dcc/LambinDF20}. 
Beyne and Verbauwhede proposed the use of algebraic transition matrices to search for generalized integral properties and applied this approach to \texttt{PRESENT}~\cite{tosc/BeyneV23}. 
\par
Artificial intelligence-based approaches to symmetric cryptanalysis, as demonstrated in works published at CRYPTO 2019~\cite{gohr2019improving}, EUROCRYPT 2021~\cite{eurocrypt/BenamiraGPT21}, and ASIACRYPT 2023~\cite{asiacrypt/bao2023more}, have shown that neural networks are capable of extracting additional non-random features from limited amounts of data. 
Consequently, a promising direction is to treat neural networks as auxiliary tools for classical cryptanalysis. Unlike classical automated methods that require significant manual effort to design and tune models, neural networks can learn novel cryptographic features directly from ciphertext datasets generated by the cipher itself. A beneficial framework is to discover novel insights from neural networks and use them to advance classical cryptanalysis methods. 
In this work, we prove the effectness of the framework by focusing on the integral properties inherent in ciphertext sets under a given plaintext structure, and improve existing integral results with what we discovered with neural networks.

\par
\begin{enumerate}
\item [-] \textbf{Improving Automated Search Models Inspired by Neural Networks.} We adopt the concept of parity sets to preprocess ciphertext multisets, ensuring that only properties relevant to integral analysis are retained. This allows the neural network to focus exclusively on learning features associated with the integral property. By comparing the integral distinguishers discovered by neural networks with those identified using classical methods under the same number of active plaintext bits, we observe that existing automated search models often fail to produce optimal distinguishers.  
To address this limitation, we propose a meet-in-the-middle strategy that combines the bit-based division property over two subsets with the monomial prediction technique—effectively mitigating the inaccuracy of the former and the computational overhead of the latter.  
As a result, our improved automated search model successfully identifies an 11-round key-independent integral distinguisher with fewer active bits for both \texttt{SKINNY-64-64} and \texttt{SKINNY-128-128}, as well as a 12-round key-dependent integral distinguisher for \texttt{SKINNY-64-64}, as shown in Table~\ref{tab:distinguisher_result}.

\vspace{0.5em}
\item [-] \textbf{Enhanced Key-Recovery Attacks Based on Integral Distinguishers.} Beyond the pursuit of longer-round distinguishers, we identify short-round distinguishers that significantly improve key-recovery attacks. In particular, we utilize a 7-round key-independent integral distinguisher to mount a 15-round key-recovery attack on \texttt{SKINNY-n-n}, by performing 3 rounds of forward decryption and 5 rounds of backward encryption—improving the previous best result by one round. Since the distinguisher is key-independent, it can also be applied to \texttt{SKINNY-n-2n} and \texttt{SKINNY-n-3n}.  
Furthermore, we employ an 8-round key-dependent integral distinguisher to reduce the complexity of key-recovery attacks on \texttt{SKINNY}. However, this attack is only effective for a subset of keys and does not apply to the full key space, as summarized in Table~\ref{tab:attack_result}.  We have made our source code publicly available at \url{https://anonymous.4open.science/r/skinny-automatic-and-AI-analysis-21BA/}..

\vspace{0.5em}
\item [-] \textbf{Neural Networks as Feature Explorers for Cryptanalysis.} Unlike classical cryptanalysis methods, which typically rely on precisely characterizing specific non-random properties, 
neural networks excel at learning features directly from the data distribution. This capability enables them to discover less well-defined features compared with classical methods.
By comparing the results of neural networks and automated search models under identical data constraints and analyzing the internal behavior of the trained models, we are able to uncover feature properties previously ignored by conventional analysis. These insights can, in turn, guide the design of more effective and delicate classical cryptanalytic techniques.  
While neural networks are inherently limited in their ability to construct long-round distinguishers due to data complexity constraints, they often outperform classical methods in the discovery of short-round distinguishers. Notably, the distinguishers used in our key-recovery attacks are directly discovered through neural networks. These findings underscore the unique and complementary role of neural networks as auxiliary tools in cryptanalysis.
\end{enumerate}
\begin{table}[ht]
\centering
\caption{The integral distinguisher against  \skinny in single tweakey setting}
\label{tab:distinguisher_result}
\SetTblrInner{colsep=1.8pt}
\begin{tblr}
{
colspec={Q[m,c]Q[m,c]Q[m,c]Q[m,c]Q[m,c]Q[m,c]},
cell{2}{1}={r=4}{},
hline{1,Z}={.08em},
hline{2}={.05em},
hline{6}={1-Z}{.05em},
}
 Cipher   & Round & Type    & {Data} & {\# Balanced \\ Bits}    & Reference \\
 \skinnyver{64}{64}   & 11     & {Key-Independent + \\ Linear Combination of Bits} & $2^{63}$  & 11  &  
 \cite{fse/derbez2020increasing} \\
 \skinnyver{64}{64}   & 11     & {Key-Independent + \\ Linear Combination of Bits} & $2^{60}$  & 16 &  Sect.\ref{sec:improved_search_model} \\
 \skinnyver{64}{64}   & 11     & {Key-Dependent +   \\ Nonlinear Combination of Bits} & $2^{60}$  & 2 &  Sect.\ref{sub:key-dependent-nonlinear-combination} \\
 \skinnyver{64}{64}   & 12     & {Key-Dependent +  \\ Linear Combination of Bits } & $2^{60}$  & 3 &  Sect.\ref{sub:key-dependent-linear-combination} \\
 \skinnyver{128}{128}   & 11     & {Key-Independent + \\ Linear Combination of Bits} & $2^{60}$  & 48 &  Sect.\ref{sec:improved_search_model} \\
\end{tblr}
\end{table}
\begin{table}[ht]
\centering
\caption{The integral attack against \skinny in single tweakey setting}
\label{tab:attack_result}
\SetTblrInner{colsep=6.0pt}
\begin{tblr}
{
colspec={Q[m,l]Q[m,c]Q[m,c]Q[m,c]Q[m,c]Q[m,c]Q[m,c]},
cell{2,9}{1}={r=3}{},
cell{5,7,12,14}{1}={r=2}{},
hline{1,Z}={.08em},
hline{2}={.05em},
hline{5,7,9,12,14,16}={1-Z}{.05em},
}
 Cipher   & Round & Configure &  {Time}        & {Data}   & {Key Space} & Reference \\
 \skinnyver{64}{64}   & 14  &  4+6+4  & $2^{48.087}$  & $2^{48}$ &  $2^{64}$    & \cite{crypto/BeierleJKL0PSSS16} \\
 \skinnyver{64}{64}   & 15  &  3+7+5  & $2^{61.976}$  & $2^{48}$ &  $2^{64}$    & Sect.\ref{15-skinny64-key-recovery-single-key} \\
 \skinnyver{64}{64}   & 15  & 3+7+5   & $2^{53.926}$  & $2^{48}$ & $2^{62}$    & Sect.\ref{15-skinny64-key-recovery-weak-key} \\
 \skinnyver{64}{128}  & 17   & 3+7+7  & $2^{126.421}$ & $2^{48}$ &  $2^{128}$   & Sect.\ref{15-skinny64-key-recovery-single-key} \\
 \skinnyver{64}{128}  & 17  & 3+7+7   & $2^{118.388}$ & $2^{48}$ & $2^{126}$   & Sect.\ref{15-skinny64-key-recovery-weak-key} \\
 \skinnyver{64}{192}  & 19   & 3+7+9  & $2^{190.695}$ & $2^{48}$ &  $2^{192}$   & Sect.\ref{15-skinny64-key-recovery-single-key} \\
 \skinnyver{64}{192}  & 19  &  3+7+9  & $2^{182.671}$ & $2^{48}$ &  $2^{190}$   & Sect.\ref{15-skinny64-key-recovery-weak-key} \\
 \skinnyver{128}{128} & 14  &  4+6+4  & $2^{96.006}$  & $2^{96}$ &  $2^{128}$   & \cite{crypto/BeierleJKL0PSSS16} \\
 \skinnyver{128}{128} & 15  & 3+7+5   & $2^{125.976}$ & $2^{96}$ &  $2^{128}$   & Sect.\ref{15-skinny64-key-recovery-single-key} \\
  \skinnyver{128}{128} & 15  & 3+7+5   & $2^{109.926}$ & $2^{96}$ &  $2^{126}$   & Sect.\ref{15-skinny64-key-recovery-weak-key} \\
 \skinnyver{128}{256} & 17  & 3+7+7   & $2^{254.421}$ & $2^{96}$ &  $2^{256}$   & Sect.\ref{15-skinny64-key-recovery-single-key} \\
 \skinnyver{128}{256} & 17  & 3+7+7   & $2^{238.388}$ & $2^{96}$ &  $2^{254}$   & Sect.\ref{15-skinny64-key-recovery-weak-key} \\
 \skinnyver{128}{384} & 19  &  3+7+9  & $2^{382.695}$ & $2^{96}$ &  $2^{384}$   & Sect.\ref{15-skinny64-key-recovery-single-key} \\
  \skinnyver{128}{384} & 19  &  3+7+9  & $2^{366.671}$ & $2^{96}$ &  $2^{382}$   & Sect.\ref{15-skinny64-key-recovery-weak-key} \\

\end{tblr}
\end{table}
The structure of the paper is as follows. Section~\ref{sec:preliminary} introduces the necessary background used throughout the paper. Section~\ref{sec:overall} presents the overall motivation and core ideas. In Section~\ref{sec:short-distinguisher}, we use neural networks to construct short-round integral distinguishers. Section~\ref{sec:more_precise_model} introduces a more precise model to better characterize the learned features. Section~\ref{sec:key-dependent-integral-distinguisher} proposes key-dependent integral distinguishers. In Section~\ref{sec:improved-attack}, we improve the key recovery attack based on the proposed techniques. Finally, Section~\ref{sec:conclusion} concludes the paper.

\section{Preliminaries}
\label{sec:preliminary}


\subsection{Brief Description of \skinny{}}\label{subSec/SKI}

\skinny is a family of tweakable block ciphers~\cite{crypto/BeierleJKL0PSSS16}. \texttt{SKINNY-n} has a $n$-bit block size. We define bit numbering from left to right. Let $c$ denote the size of a cell in \skinny.
The state is represented as a $4 \times 4$ array, where each cell has a size of $c$ bits.
The input state of $r$th round is denoted by
\begin{equation*}
	\textbf{S}_r = \left(
	\begin{array}{cccc}
		\textbf{S}_r[0] & \textbf{S}_r[1] &\textbf{S}_r[2]& \textbf{S}_r[3]\\
		\textbf{S}_r[4] &\textbf{S}_r[5]&\textbf{S}_r[6]&\textbf{S}_r[7] \\
		\textbf{S}_r[8] & \textbf{S}_r[9] & \textbf{S}_r[10] & \textbf{S}_r[11]\\
		\textbf{S}_r[12] & \textbf{S}_r[13] &\textbf{S}_r[14] & \textbf{S}_r[15]
	\end{array}
	\right).
\end{equation*}

\begin{figure}[ht]
    \centering
    \includegraphics[width=1.0\textwidth]{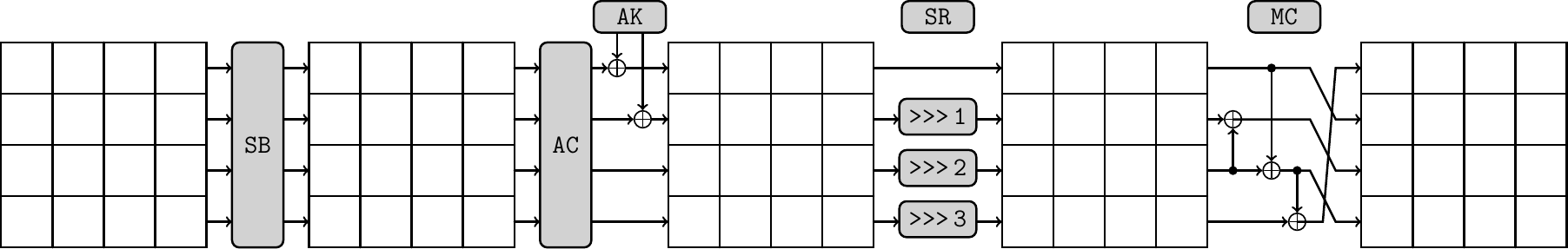}
    \caption{The round function of \skinny}
    \label{fig:skinny_round_number}
\end{figure}

For each block size $n$, \skinny-$n$ supports three tweakey sizes: $t = n$, $t = 2n$, and $t = 3n$. \skinny-$n$-$t$ denotes the version with block size $n$ and tweakey size $t$. The $t$-bit tweakey is arranged as a set of $t/n$ $4 \times 4$ arrays, denoted by $TK_z$, where $z \in \{1, \dots, t/n\}$.
Each tweakey undergoes a permutation at the beginning of each round. Additionally, every cell in the first and second rows of $TK_2$ and $TK_3$ is individually updated using a linear feedback shift register (LFSR), with the details of the LFSR described in \cite{crypto/BeierleJKL0PSSS16}.
The round function consists of five operations: SubCells (SB), AddConstants (AC), AddRoundTweakey (AK), ShiftRows (SR), and MixColumns (MC), as illustrated in Fig.~\ref{fig:skinny_round_number}.
\par

SB substitutes each cell with a $c$-bit S-box. For \skinnyver{64}{64} (resp. \skinny \allowbreak \texttt{-128-128}), $c$=4 (resp. 8).
AC updates the state by XORing it with round constants.
AK updates the state by XORing the first two rows of the state with the $t/n$ tweakey arrays.
SR rotates the $i$-th row to the right by $i$ cells.
MC multiplies each column of the state by a binary matrix:
\begin{equation*}
	\left(
	\begin{array}{cccc}
		1 & 0 &1 & 1\\
		1 & 0 & 0 &0 \\
		0 & 1 & 1 &0 \\
		1 & 0 & 1 &0
	\end{array}
	\right).
\end{equation*}

\subsection{Integral Property}
\noindent \textbf{Notation}  Let $\mathbb{F}_2$ denote the finite field $\{0,1\}$, and let $a = (a_0, a_1, \ldots, a_{m-1}) \in \mathbb{F}_2^m$ represent an $m$-bit vector, where $a[i]$ denotes the $i$-th bit of $a$. The Hamming weight $wt(a)$ is defined as
$wt(a) = \sum_{i=0}^{m-1} a[i]$.
For any $v \in \mathbb{F}_2^m$ and $v' \in \mathbb{F}_2^m$, we denote $v \succeq v'$ if $v_i \ge v_i'$ holds for all $i = 0, 1, \ldots, m-1$.
\par
The integral property utilises a set of chosen plaintexts in which certain bits take all possible values while the remaining bits are fixed to a constant. The corresponding ciphertexts are then computed using an encryption oracle. If the XOR of all ciphertexts in the set always results in 0, the cypher is said to possess an integral distinguisher.
The division property, originally proposed in \cite{conf/eurocrypt/Todo15}, provides a more precise and generalized method for identifying integral distinguishers.
\par
\vspace{0.5em}
\noindent \textbf{Bit Product Functions $\pi_u$
\cite{conf/eurocrypt/Todo15}.} Let $\pi_u : \mathbb{F}_2^m \rightarrow \mathbb{F}_2$ be a function for any $u \in \mathbb{F}_2^m$. Let $x \in \mathbb{F}_2^m$ be an input of $\pi_u$, and $\pi_u(x)$ is the AND of $x[i]$ satisfying $u[i] = 1$, namely, it is defined as
\[ \pi_u(x) := \prod_{i=0}^{m-1} x[i]^{u[i]}. \]
\par
\begin{definition}[Bit-based Division Property using Two Subsets~\cite{conf/fse/TodoM16}]
A set $\mathbb{X} \in \mathbb{F}_2^{m}$ has division property $D_{\mathbb{K}}^m$, where $\mathbb{K} \in \mathbb{F}_2^m$ is a set, if for all $u \in \mathbb{F}_2^m$, we have
\[
    \bigoplus_{x \in \mathbb{X}}\pi_{u}(x) = \begin{cases}unknown & \text    { if there is } k \in \mathbb{K} \   s.t. \  u \succeq k,  \\ 0 &     \text { otherwise. } \end{cases}
\]
\end{definition}

\begin{definition}[Algebraic Normal Form]
    The Algebraic Normal Form (ANF) of a Boolean function $f:\mathbb{F}_2^m\rightarrow\mathbb{F}_2$ can be expressed as
    $$
    f(x)=f(x_0,\cdots,x_{m-1})=\underset{u\in\mathbb{F}_2^m}{\bigoplus}a_{u}\pi_{u}(x),
    $$
    where $a_{u}\in\mathbb{F}_2$ and $\pi_{u}(x)$ is called a monomial. If the coefficient of $\pi_{u}(x)$ in $f$ is $1$, we say $\pi_{u}(x)$ is contained in $f$, denoted by $\pi_{u}(x)\rightarrow f$.
\end{definition}

\begin{definition}[Monomial Prediction~\cite{asiacrypt/HuSW020}]
If there exists a monomial sequence satisfied
$$
\pi_{u^{(0)}}(x^{(0)})\rightarrow\pi_{u^{(1)}}(x^{(1)})\rightarrow\cdots\rightarrow\pi_{u^{(r)}}(x^{(r)}),
$$
we call that there is a monomial trail connecting $\pi_{u^{(0)}}(x^{(0)})$ and $\pi_{u^{(r)}}(x^{(r)})$, denoted by $\pi_{u^{(0)}}(x^{(0)})\leadsto\pi_{u^{(r)}}(x^{(r)}).$ If the number of monomial trails connecting $\pi_{u^{(0)}}(x^{(0)})$ and $\pi_{u^{(r)}}(x^{(r)})$ is odd, then we have $\pi_{u^{(0)}}(x^{(0)})\rightarrow\pi_{u^{(r)}}(x^{(r)}).$
\end{definition}
\par
 Beyne and Verbauwhede\cite{tosc/BeyneV23} consider a general definition of integral properties that encompasses the original integral properties from \cite{conf/fse/KnudsenW02}, division properties\cite{conf/eurocrypt/Todo15,conf/fse/TodoM16} and the properties from the linearly equivalent S-boxes method of \cite{dcc/LambinDF20}, which was further developed by Derbez and Fouque~\cite{fse/derbez2020increasing}.
\begin{definition}[Integral Property~\cite{tosc/BeyneV23}\label{def:generalized-integral-property}]
Let $F:\mathbb{F}_2^n\rightarrow \mathbb{F}_2^m$ be a vectorial Boolean function. An integral property for $F$ is a pair ($\mathbb{X},g$) with $\mathbb{X}\in\mathbb{F}_2^n$ and $g:\mathbb{F}_2^m\rightarrow \mathbb{F}_2$, and its evaluation is equal to
\begin{equation}
\label{equation:generalized-integral-property}
   \underset{x\in \mathbb{X}}{\sum}g(F(x)).
\end{equation}

\end{definition}

\section{The Overall Motivation and Core Ideas}
\label{sec:overall}
In cryptanalysis, distinguishers are constructed based on non-random features observed in plaintext-ciphertext pairs to distinguish ciphertexts from random values. Meanwhile, neural networks are data-driven models capable of classifying inputs by learning discriminative patterns from labeled data. This naturally motivates the idea of integrating neural networks into cryptanalytic methodologies. Recent studies, including CRYPTO~2019~\cite{gohr2019improving}, EUROCRYPT~2021~\cite{eurocrypt/BenamiraGPT21}, and ASIACRYPT~2023~\cite{asiacrypt/bao2023more}, have demonstrated that neural networks can uncover previously unexplored features, offering novel insights into modern cryptanalysis. Therefore, we propose employing neural networks as an auxiliary tool in integral cryptanalysis.

High-round distinguishers typically require significantly more data, which inherently limits the ability of neural networks to distinguish between ciphertexts and random values in such settings. Consequently, our objective is not to construct high-round distinguishers using neural networks, but rather to extract richer and more informative features from limited data.

Since the features learned by neural networks are entirely derived from data, and interpretability is essential in cryptanalysis, we introduce the concept of parity sets in Section~\ref{sub:define_data_type} to constrain the type of features embedded in the training instances. By encoding input data as vectorial division sequences, we successfully obtain 7-round and 8-round integral distinguishers with low data complexity, as presented in Section~\ref{sub:linear-combiniation-bit-distinguisher}.

In Section~\ref{sec:compare-neural-and-classical}, we compare the distinguishers discovered by neural networks with those found by existing automated search models. This comparison reveals that the latter are not always precise: for a fixed number of active bits, automated approaches may fail to identify optimal distinguishers. Motivated by this observation, we enhance the automated search framework by incorporating the division property with two subsets and the monomial prediction technique. The resulting improved model not only reproduces the distinguishers found by neural networks but also yields longer distinguishers requiring fewer active bits than prior results, as detailed in Section~\ref{sec:improved_search_model}.

Furthermore, in Section~\ref{sub:key-depent-linear-combination}, we observe that neural networks are capable of learning both deterministic and probabilistic integral features. Our analysis shows that the probabilistic nature of some features stems from key-dependent monomials, where variations in key values influence the presence of integral properties. Leveraging this insight, we extend our search strategy and identify a 12-round key-dependent integral distinguisher for \texttt{SKINNY-64-64}, which, to the best of our knowledge, represents the longest known distinguisher for this cipher (see Section~\ref{sub:key-dependent-linear-combination}). Most importantly, guided by neural network, we progressively expand the feature space in our search model—from linear combinations of bits to nonlinear combinations—as described in Section~\ref{sub:key-dependent-nonlinear-combination}.

While our improved model is designed to identify long-round distinguishers, we emphasize that longer distinguishers do not necessarily lead to better key-recovery attacks. In Section~\ref{15-skinny64-key-recovery-single-key}, we demonstrate a 15-round key-recovery attack on \texttt{SKINNY-64-64} using a 7-round distinguisher discovered by a neural network, combined with a structure that performs 3 rounds of forward decryption and 5 rounds of backward encryption. This surpasses the previous best result by one round. Moreover, in Section~\ref{15-skinny64-key-recovery-weak-key}, we show that using probabilistic integral distinguishers can further reduce the time complexity of the attack. However, this particular attack is not valid for all key values.

In summary, neural networks serve as a valuable auxiliary tool in cryptanalysis. Although the features in our training data are constrained to those related to integral properties—limiting the discovery of entirely new patterns—neural networks effectively reveal deficiencies in existing automated search models. By addressing these deficiencies, we achieve stronger cryptanalytic results. To a certain extent, this demonstrates that neural networks can significantly contribute to the advancement of classical cryptanalytic techniques.

\section{Construct Short-Round Integral Distinguisher Using Neural Network}
\label{sec:short-distinguisher}
Neural networks have the potential to learn additional non-random features from ciphertexts beyond what can be extracted by classical cryptanalysis methods. In CRYPTO 2019, Gohr used ciphertext pairs generated from plaintext pairs with a fixed input difference as inputs to a neural network, enabling it to learn difference-related features~\cite{gohr2019improving}. Bao \etal{} further demonstrated that neural networks not only learn differential features from ciphertext pairs of \speck, but also extract XOR-related information between the left and right branches of the ciphertext~\cite{asiacrypt/bao2023more}. Inspired by these findings, we aim to explore whether neural networks can also discover novel integral properties directly from data.

\subsection{Definition of Data Types}
\label{sub:define_data_type}
Zahednejad \etal{} used multisets as inputs to a neural network, training it to distinguish between ciphertexts with specific integral properties and random values~\cite{ijis/ZahednejadL22}. The positive instances are obtained by encrypting plaintext multisets generated through partial-bit enumeration, while the negative instances are generated by encrypting randomly constructed plaintext multisets. However, training integral distinguishers using neural networks presents several limitations:
\begin{enumerate}

    \item In classical integral cryptanalysis, extending the length of an integral distinguisher typically requires activating a large number of plaintext bits. However, encrypting such a lot of plaintexts is computationally expensive. Consequently, when training an integral distinguisher using a neural network, only a small number of plaintext bits can be activated, which limits the neural distinguisher to fewer rounds.
    \vspace{0.5em}
    \item Since the multisets consist of concrete ciphertext values, they may contain various types of non-random features. Neural networks may unintentionally learn to exploit a mixture of these features during training, whereas the original goal is to guide the network to focus exclusively on integral-related properties.
\end{enumerate}
\par
The data requirements described in Limitation~1 are inherent to cryptanalysis in general—not only integral cryptanalysis, but also differential and linear cryptanalysis face similar challenges. The motivation for introducing neural networks into cryptanalysis is precisely to uncover more non-random features from a limited amount of data.
We aim to address Limitation~2 by exploring methods to reduce the influence of unintended non-random features.
In CRYPTO 2016, Boura \etal{} proposed the notion of the parity set to characterize the division property of a ciphertext set.
\begin{definition}[\textbf{Parity Set}~\cite{crypto/BouraC16}]
Let $\mathbb{X}$ be a set of elements in $\mathbb{F}_2^m$, the parity set of $\mathbb{X}$, denoted by $\mathcal{U}(\mathbb{X})$, is the subset of $\mathbb{F}_2^m$ defined by
$$\mathcal{U}(\mathbb{X}) = \{u \in \mathbb{F}_2^m: \bigoplus_{x\in \mathbb{X}} x^u=1\}.$$
\end{definition}
\par
We propose leveraging the concept of the parity set to generate training data that is better aligned with the learning of integral-related features using neural networks. In practice, it is generally infeasible to characterize the division property of the entire cipher state in a single step. Instead, the analysis is typically performed at the granularity of individual S-boxes. Consequently, the division property of the full state can be derived by aggregating the division properties of each individual cell.
Therefore, we define two types of data formats as follows.
\begin{definition}[\textbf{Division Suquence}]\label{def:division_sequence}
Let $\mathbb{X}$ be a set of elements in $\mathbb{F}_2^c$, the division sequence of $\mathbb{X}$, denoted by $\mathbb{DS}$, is a sequence defined by
$$\mathbb{DS} = [\bigoplus_{x\in \mathbb{X}} x^u,u \in \mathbb{F}_2^c].$$
\end{definition}
\par
\begin{definition}[\textbf{Vectorial Division Suquence}]\label{def:vectorial_division_sequence}
Let $\mathbb{X}_1,\mathbb{X}_2,\cdots \mathbb{X}_{i}\ (1\le i \le \frac{n}{c})$ be the set of elements in $\mathbb{F}_2^c$, the vectorial division sequence of $\mathbb{X}_1,\mathbb{X}_2,\cdots \mathbb{X}_{s}$, denoted by $\mathbb{VDS}$, is a sequence defined by
$$\mathbb{VDS} = [\bigoplus_{x\in \mathbb{X}_1}x^u] \parallel [\bigoplus_{x\in \mathbb{X}_2}x^u] \parallel \cdots \parallel [\bigoplus_{x\in \mathbb{X}_i}x^u] ,u \in \mathbb{F}_2^c.$$
\end{definition}
\par
We use the division sequence $\mathbb{DS}$ to describe the integral property of a single cell, and extend this notion to the vectorial division sequence $\mathbb{VDS}$ to represent the collective integral properties of multiple cells.
\subsection{The Neural Network Learns the Integral-Related Property}
\label{subsec:about-neural-network}
The following procedure enables the neural network to learn integral-related properties by using the (vectorial) division sequence as input.

\par
\vspace{0.5em}
\noindent \textbf{Data Generation}  Let $N$ and $M$ denote the sizes of the training set and the test set, respectively. The dataset is generated through the following procedure:
\begin{enumerate}
    \item[-] Label generation. Generate $N$ (resp., $M$) labels $Y$ for the $N$ (resp., $M$) instances, with approximately half labeled as 0 (i.e., negative instances) and the other half as 1 (i.e., positive instances).
    \vspace{0.5em}

    \item[-] Plaintext multiset generation. Use the Linux random number generator to generate $N$ (resp., $M$) uniformly distributed plaintexts $P_i$. Then, obtain a plaintext multiset by traversing a subset of plaintext bits while keeping the remaining bits fixed.
    \vspace{0.5em}

    \item[-] Ciphertext multiset generation. If the label $Y$ is 0, replace the traversed plaintext bits with random values. Use the Linux random number generator to generate $N$ (resp., $M$) uniformly distributed plaintexts $K_i$. Encrypt the resulting plaintext multiset using the corresponding key $K_i$ to obtain the ciphertext multiset.
    \vspace{0.5em}

    \item[-] Instance formatting. Construct the input instance from the ciphertext multiset according to Definition~\ref{def:division_sequence} or Definition~\ref{def:vectorial_division_sequence}.
\end{enumerate}

\par
\vspace{0.5em}
\noindent \textbf{Network Architecture}
We employ a simple three-layer fully connected neural network. The three layers contain 256, 64, and 1 neurons, respectively. Each fully connected layer is followed by a batch normalization layer and a ReLU activation function. Finally, a sigmoid activation function is applied to produce a scalar output in the range $[0, 1]$. If the output is greater than 0.5, the input is classified as a positive instance; otherwise, it is classified as a negative instance.
\par
\vspace{0.5em}
\noindent \textbf{Training Process}
The neural network was trained for 20 epochs using a training dataset of size $N$ and a test dataset of size $M$. The batch size was dynamically adjusted to optimize GPU performance. Optimization was performed using the Adam algorithm with the mean square error (MSE) as the loss function.
To improve training efficacy, a cyclic learning rate schedule was adopted, defined as:
$
l_{i} = \alpha + \frac{(9 - i) \bmod 10}{10} \cdot (\beta - \alpha),
$
where $\alpha = 10^{-4}$ and $\beta = 2 \times 10^{-3}$. After each epoch, the model was saved, and the best-performing network was selected based on validation loss for subsequent evaluation on the test set.

\par
\vspace{0.5em}
\noindent \textbf{Performance Evaluation of the Integral-Neural Distinguisher}
Accuracy (Acc) serves as the primary metric for evaluating the performance of the integral-neural distinguisher. In addition, the true positive rate (TPR) and true negative rate (TNR) are reported to assess the proportions of correctly classified positive and negative instances, respectively.

\subsection{Integral Distinguisher Based on Linear Combination of Bits}
\label{sub:linear-combiniation-bit-distinguisher}
In this section, the neural network learns the integral properties in Definition~\ref{def:generalized-integral-property} where the function $g$ is linear. As a result, we obtain extended integral distinguishers under a given plaintext structure previously proposed by Lambin \etal{} \cite{dcc/LambinDF20} and further developed by Derbez \etal{} \cite{fse/derbez2020increasing}, which we refer to as integral distinguisher based on linear combination of bits.
\par
We use the vectorial division sequence $\mathbb{VDS}$ as the input of the neural network to train an integral-neural distinguisher. The size of $\mathbb{VDS}$ is 256, as $u \in \mathbb{F}_2^4$ and there are 16 cells in \skinnyver{64}{64}. To account for the computational cost of data generation, the sizes of the training and test sets, $N$ and $M$, are both set to $10^6$. The network architecture and training process are detailed in Section~\ref{subsec:about-neural-network}. The result of the 7-round integral-neural distinguisher for \skinnyver{64}{64} is shown in Table~\ref{tab:7-round-64-15-cell-distinguisher}.
\begin{table}[htb]
\centering
\caption{The integral-neural distinguisher of 7-round \skinnyver{64}{64} using $\mathbb{VDS}$}
\label{tab:7-round-64-15-cell-distinguisher}
\begin{tblr}
{
colspec={Q[m,c]Q[m,c]Q[m,c]Q[m,c]},
hline{1,Z}={.08em},
hline{2}={.05em},
cell{2}{3}={gray9},
}
 {Activated Plaintext Cell} & Acc   & TPR   & TNR \\
15th             & 99.8\% & 100\% & 99.6\%  & \\
\end{tblr}
\end{table}
\par
\vspace{0.5em}
\noindent \textbf{Observation of Performance Metrics} The true positive rate (TPR) reaches 100\%, indicating that the neural network has learned deterministic features, as all positive instances conform to these features. For negative instances, there is a $1 - 0.096 = 2^{-8}$ probability that a negative instance is incorrectly classified as positive. Since negative instances make up half of the dataset, the integral-neural distinguisher misclassifies approximately $2^{-9}$ of the total dataset, which corresponds to an overall error rate of 0.2\% and an accuracy of 99.8\%.
In other words, the feature learned by the 7-round integral-neural distinguisher is associated with only 8 bits.
\par
\begin{figure}[htb]
    \centering
    \includegraphics[width=0.6\textwidth]{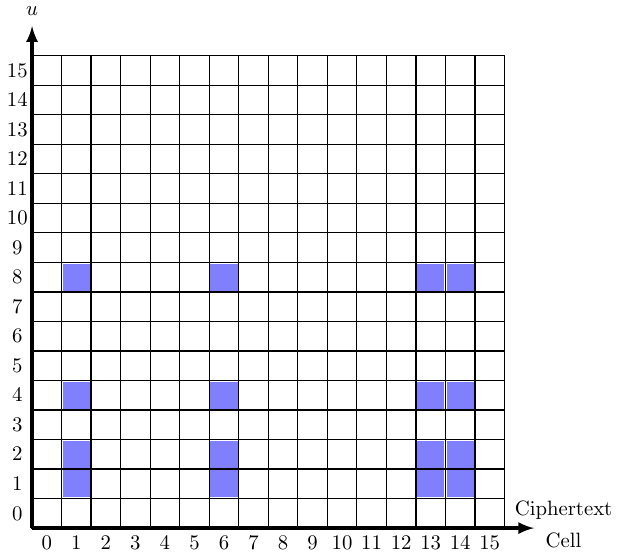}
    \caption{The bit sensitivity test for the 7-round integral-neural distinguisher}
    \label{fig:bit-sen-result-7-round-vectorial-division}
\end{figure}

\vspace{0.5em}
\noindent \textbf{The Working Mechanism of the Integral-Neural Distinguisher}
Once an integral-neural distinguisher is trained, the decision rules it relies on are implicitly fixed. By modifying the input, we can probe the underlying patterns learned by the neural network. Yi Chen \etal{}~\cite{cj/ChenSY23} proposed an algorithm, called the \textit{bit sensitivity test}, to evaluate the impact of individual bits in a given instance on the performance of the neural distinguisher, as outlined in Algorithm~\ref{alg:bit_sensitively_test}.
We apply the bit sensitivity test to identify which specific bits significantly influence the accuracy of the integral-neural distinguisher. The results are shown in Fig.~\ref{fig:bit-sen-result-7-round-vectorial-division}. The horizontal axis represents the ciphertext cell corresponding to each bit in the $\mathbb{VDS}$, while the vertical axis indicates the value of $u$ associated with that bit. The \fillcolor{blue!50} shading highlights the bits for which randomization causes a significant drop in the distinguisher's accuracy.

\par

From Fig.~\ref{fig:bit-sen-result-7-round-vectorial-division}, it is evident that the 7-round integral-neural distinguisher learns features from 16 bits of the $\mathbb{VDS}$. However, based on the performance metrics, the feature actually utilized by the neural network appears to be related to only 8 bits. To resolve this discrepancy, we extract the values of these 16 bits and observe several interesting phenomena.
Specifically, the 4 bits corresponding to the four different $u$ values on ciphertext cells $\{1, 13\}$ are identical. In other words, the 4-bit XOR sum over ciphertext cells $\{1, 13\}$ is always \texttt{0b0000}. A similar pattern is observed for ciphertext cells $\{6, 14\}$.
Therefore, it can be deduced that the integral-neural distinguisher classifies instances based on this 8-bit pattern. The working mechanism of the 7-round integral-neural distinguisher can thus be summarized as follows:
\par
\begin{pattern}
\label{7-round-skinny-64-64-pattern}
    For 7-round \skinnyver{64}{64}, when the 15th plaintext cell is activated, then $\bigoplus\limits_{x_1\in \mathbb{X}_1} \pi_u(x_1) \ \oplus \bigoplus\limits_{x_2\in \mathbb{X}_2} \pi_u(x_2) = 0 $, where $wt(u)=1$ and ciphertext cells $(\mathbb{X}_1,\mathbb{X}_2) \in \{(1,13),(6,14)\}$.
\end{pattern}
\par
Moreover, the Pattern also applies to \textsc{SKINNY}-\texttt{128}-\texttt{128}.
We convert Pattern~\ref{7-round-skinny-64-64-pattern} into the representation of a integral distinguisher.

\par
\vspace{0.5em}
\begin{distinguisher}\label{dis:integral-distinguisher-7-round}
\emph{\textbf{(Key-Independent Integral Distinguisher Based on Linear Combination of Bits Against 7-round \texttt{SKINNY-n-n})}.}
When the 15th plaintext cell is activated, the linear combination of bits $b_i \oplus b_{i+12c},i\in \{4,5,6,7\}$ and $b_i \oplus b_{i+8c},i\in \{24,25,26,27\}$ are balanced, where c is the size of the S-box.
\end{distinguisher}
\par
Using the same training method and analysis procedure, we obtained an 8-round integral distinguisher based on a linear combination of bits by activating two plaintext cells.
\begin{distinguisher}\label{dis:integral-distinguisher-8-round}
\emph{\textbf{(Key-Independent Integral distinguishers Based on Linear Combination of Bits Against 8-round \skinnyver{64}{64}.)}}
If 14th and 15th plaintext cells are activated, then the linear combination of bits $b_{28}\oplus b_{44}\oplus b_{60}$ is balanced.
\end{distinguisher}

\subsection{Transforming Neural Distinguisher to Boolean Function}
\label{sub:transform-distinguisher-to-boolean-function}
We introduce an alternative approach to analyzing the working mechanism of the integral-neural distinguisher. This method is introduced here in preparation for its application in the subsequent sections. The working mechanism of the 7- (or 8-) round integral-neural distinguisher can be understood by directly observing the values of the sensitive bits, primarily because the underlying rules it relies on are relatively simple.
\par
Since both the input and output of the neural network can be interpreted as binary values, the integral-neural distinguisher can be transformed into a Boolean function. By deriving its conjunctive normal form (CNF), we can precisely characterize the rules learned by the neural network. The transformation process is described in Algorithm~\ref{algorithm:cnf-expression}, as shown below.
\begin{algorithm}[htp]
\caption{Get the CNF of the Neural Distinguisher}\label{algorithm:cnf-expression}
\begin{algorithmic}[1]
\renewcommand{\algorithmicrequire}{\textbf{Input:}}
\renewcommand{\algorithmicensure}{\textbf{Output:}}
\Require a trained neural distinguisher $\mathcal{ND}$, the size of the input to the distinguisher $d$
\Ensure The CNF of integral-nerual distinguisher $\text{CNF}_\mathcal{ND}$
\State $T_t \leftarrow [ \ ] $ \Comment{Truth table}
\For {input in 0 to $2^d$}
\If {$\mathcal{ND}(\text{input})>0.5$}
\State add 1 to $T_t$
\Else
\State add 0 to $T_t$
\EndIf
\EndFor
\State $\mathcal{BF} $ = BooleanFunction($T_t$) \Comment{Construct a boolean function}
\State Get $\text{CNF}_\mathcal{ND}$ of  $\mathcal{BF} $
\State \Return $\text{CNF}_\mathcal{ND}$
\end{algorithmic}
\end{algorithm}
\par
\vspace{0.5em}
\noindent \textbf{Explaining the Working Mechanism of the Integral-Neural Distinguisher}
The size of the $\mathbb{VDS}$ is 256, which makes it infeasible to construct a Boolean function over $\mathbb{F}_2^{256}$. Moreover, according to the results shown in Fig.~\ref{fig:bit-sen-result-7-round-vectorial-division}, the features learned by the neural network are related to only 16 bits. Therefore, we retrain the integral-neural distinguisher using these 16 sensitive bits (indexed as $\bm{b}_{cth}^u$, where $cth$ denotes the index of the cell), and transform the resulting model into a Boolean function, along with its CNF, using Algorithm~\ref{algorithm:cnf-expression}.
We then apply the \texttt{Espresso} logic minimizer\footnote{\url{https://github.com/classabbyamp/espresso-logic}} to obtain the minimized CNF corresponding to the 7-round integral-neural distinguisher, which is shown below:
\\
$(\neg \bm{b}_{1}^{1} \lor \bm{b}_{13}^{1} )\And (\bm{b}_{1}^{1} \lor \neg \bm{b}_{13}^{1}   )\And
(\neg \bm{b}_{1}^{2} \lor \bm{b}_{13}^{2})\And (\bm{b}_{1}^{2} \lor \neg \bm{b}_{13}^{2})\And
(\neg \bm{b}_{1}^{4} \lor \bm{b}_{13}^{4})\And (\bm{b}_{1}^{4} \lor \neg \bm{b}_{13}^{4})\And
(\neg \bm{b}_{1}^{8} \lor \bm{b}_{13}^{8})\And (\bm{b}_{1}^{8} \lor \neg \bm{b}_{13}^{8})\And
(\neg \bm{b}_{6}^{1} \lor \bm{b}_{14}^{1})\And (\bm{b}_{6}^{1} \lor \neg \bm{b}_{14}^{1})\And
(\neg \bm{b}_{6}^{2} \lor \bm{b}_{14}^{2})\And (\bm{b}_{6}^{2} \lor \neg \bm{b}_{14}^{2})\And
(\neg \bm{b}_{6}^{4} \lor \bm{b}_{14}^{4})\And (\bm{b}_{6}^{4} \lor \neg \bm{b}_{14}^{4})\And
(\neg \bm{b}_{6}^{8} \lor \bm{b}_{14}^{8})\And (\bm{b}_{6}^{8} \lor \neg \bm{b}_{14}^{8})$
\par
To make the CNF evaluate to true, all clauses must be satisfied, which implies that the ciphertext must satisfy every clause. Based on this, we infer that the following expressions must all evaluate to 0:
\[
\bm{b}_{1}^{1} \oplus \bm{b}_{13}^{1},\quad
\bm{b}_{1}^{2} \oplus \bm{b}_{13}^{2},\quad
\bm{b}_{1}^{4} \oplus \bm{b}_{13}^{4},\quad
\bm{b}_{1}^{8} \oplus \bm{b}_{13}^{8},\quad
\bm{b}_{6}^{1} \oplus \bm{b}_{14}^{1},\quad
\bm{b}_{6}^{2} \oplus \bm{b}_{14}^{2},\quad
\bm{b}_{6}^{4} \oplus \bm{b}_{14}^{4},\quad
\bm{b}_{6}^{8} \oplus \bm{b}_{14}^{8}.
\]
This result is consistent with the observed pattern in Pattern~\ref{7-round-skinny-64-64-pattern}, which clearly demonstrates the effectiveness of the proposed method.

\section{More Precise Automated Search Model} 
\label{sec:more_precise_model}
In Sect.~\ref{sec:compare-neural-and-classical}, by comparing existing classical results with those obtained from integral-neural distinguishers, we observe that, for the same number of rounds, integral distinguishers found by existing automated methods typically require more active plaintext bits than those identified by neural networks. This indicates that current automated modeling techniques may lack precision and fail to identify optimal distinguishers. 
Motivated by this observation, we refine the automated search model to align its results with those of the neural approach. With the improved model, we are able to discover more effective integral distinguishers in Sect.~\ref{sec:improved_search_model}.

\subsection{Comparison of Integral Distinguishers Based on Neural Networks and Classical Methods}
\label{sec:compare-neural-and-classical}
In CRYPTO 2016, Beierle \etal{} proposed a 6-round integral distinguisher for both \skinnyver{64}{64} and \skinnyver{128}{128} by activating one plaintext cell~\cite{crypto/BeierleJKL0PSSS16}. In 2019, Wening Zhang \etal{} discovered a 7-round integral distinguisher for \textsc{SKINNY} by activating two plaintext cells~\cite{tosc/ZhangCGP19}. At FSE 2020, Derbez \etal{} obtained an 8-round integral distinguisher based on linear combinations of bits, using 15 active plaintext bits, by employing the Superbox-Sbox technique and linear mappings~\cite{fse/derbez2020increasing}. 

\begin{table}[H]
\centering
\caption{The integral distinguisher against  \skinnyver{64}{64}}
\label{tab:linear-distinguisher_result}
\SetTblrInner{colsep=11.5pt}
\begin{tblr}
{
colspec={Q[m,c]Q[m,c]Q[m,c]Q[m,c]Q[m,c]},
cell{3}{1}={r=2}{},
cell{5}{1}={r=2}{},
hline{1,Z}={.08em},
hline{2}={.05em},
hline{3,5}={1-Z}{.05em},
}
Round & Method & Type    & {Data}     & Reference \\
6     & Classical & Single Ciphertext Bit & $2^{4}$  & \cite{crypto/BeierleJKL0PSSS16} \\
7     & Classical & Single Ciphertext Bit & $2^{8}$  & \cite{tosc/ZhangCGP19} \\
7     & Neural & { Linear Combination of Bits} & $2^{4}$  & Sect.\ref{sub:linear-combiniation-bit-distinguisher} \\
8     & Classical & {Linear Combination of Bits} & $2^{15}$  & \cite{fse/derbez2020increasing} \\
8     & Neural & {Linear Combination of Bits} & $2^{8}$  & Sect.\ref{sub:linear-combiniation-bit-distinguisher} \\
\end{tblr}
\end{table}
\par
Based on the results in Table~\ref{tab:linear-distinguisher_result}, we present the following observations and conjectures:
\begin{itemize}
    \item According to~\cite{crypto/BeierleJKL0PSSS16} and Distinguisher~\ref{dis:integral-distinguisher-8-round} in Section~\ref{sub:linear-combiniation-bit-distinguisher}, the use of different types of distinguishers can increase the number of rounds achievable by integral distinguishers under the same number of activated plaintext bits.
    \vspace{0.5em}
    \item By comparing the integral distinguisher in~\cite{fse/derbez2020increasing} with Distinguisher~\ref{dis:integral-distinguisher-8-round} in Section~\ref{sub:linear-combiniation-bit-distinguisher}, both of which are based on linear combinations of bits, we observe a difference in the number of required active plaintext bits. 
\end{itemize}
These discrepancy may be attributed to the lack of precision in the existing automated modeling method.

\subsection{Improved Automated Search Model}
\label{sec:improved_search_model}
\par
We attempt to reproduce result in Table~\ref{tab:linear-distinguisher_result} using existing modeling methods:
\begin{itemize}
    \item \textbf{Bit-Based Division Property with Two Subsets}~\cite{asiacrypt/xiang2016applying}: Using this method, we are only able to find a 6-round integral distinguisher when activating 4 plaintext bits.
    \vspace{0.5em}
    \item \textbf{Monomial Prediction Technique}~\cite{asiacrypt/HuSW020}: This method fails due to the high algebraic degree and the numerous copy operations in the \texttt{SKINNY} model. The number of candidate monomials becomes so large that it is extremely difficult to determine whether a specific monomial appears in the ANF.
\end{itemize}
\par
The automated search model based on the bit-based division property with two subsets is simple but lacks precision. In contrast, monomial prediction techniques offer high precision but suffer from high computational complexity. To balance these trade-offs, we developed an automated search model based on the meet-in-the-middle approach. This method  applies backward monomial extension for $q$ rounds and performs forward modeling using the bit-based division property with two subsets for $p$ rounds, thereby enabling the search for a $(p+q)$-round integral distinguisher. The core idea of the improved automated search model is illustrated as follows:
\[
\underbrace{\text{Input division property } \bm{d}_0 \rightarrow \mathbb{D}_p}_{p\text{ rounds forward modeling}} \rightarrow \textbf{Check} \leftarrow \underbrace{\text{Monomials} \leftarrow \text{Ciphertext bit}}_{q\text{ rounds backward extension}}.
\]
\par
\vspace{0.5em}
\noindent \textbf{Notations} 
Let $\bm{x} = (x_0, \cdots, x_{63})$, $\bm{s} = (s_0, \cdots, s_{63})$, $\bm{b} = (b_0, \cdots, b_{63})$, and $\bm{k} = (k_0, \cdots, k_{63})$ denote the plaintext, the output after $p$ rounds, the $p+q$ rounds ciphertext, and the key of \skinnyver{64}{64}, respectively where bits are indexed from left to right. The encryption from round $r_i$ to $r_j$ is denoted as $E_{\bm{k}}^{(r_i, r_j)}$, so $\bm{b} = E_{\bm{k}}^{(0, p+q)}(\bm{x}) = E_{\bm{k}}^{(p, p+q)}(\bm{s})$. The input division property is denoted by $\bm{d}_0$, and the set of division properties after $p$ rounds is denoted as $\mathbb{D}_{p}$.

\par

\par
The MILP-based automated modeling of division properties is already well-established in \cite{asiacrypt/xiang2016applying} and will not be elaborated here. The basic procedure of the improved automated search model is presented as follows:
\begin{enumerate}
    \item[-] \textbf{$q$-Round Backward Monomial Extension.} For the linear combination of bits \(\mathcal{C}(\bm{b})\), we recursively expand it in reverse to its ANF over the $p$-round intermediate variables \(\bm{s}\):
    \[
    \mathcal{C}(\bm{b}) = \mathcal{C}(E_{\bm{k}}^{(0, p+q)}(\bm{x})) = \mathcal{C}(E_{\bm{k}}^{(p, p+q)}(\bm{s})) = \sum_{i} \pi_{\bm{u}_i}(\bm{s}) \, \pi_{\bm{v}_i}(\bm{k}).
    \]
    We focus only on the balance property of each \(\pi_{\bm{u}_i}(\bm{s})\), treating \(\pi_{\bm{v}_i}(\bm{k})\) as constants. To accelerate the search, we apply Reducing Rule~\ref{reducing-rule-1} from Lambin's work~\cite{dcc/LambinDF20} to reduce the number of monomials that need to be examined.
    \vspace{0.5em}

    \item[-] \textbf{$p$-Round Forward Modeling Using the Bit-Based Division Property with Two Subsets.} We model the propagation of the division property over $p$ rounds, storing all possible division properties in \(\mathbb{D}_{p}\). For details on the modeling process of the bit-based division property with two subsets, please refer to~\cite{asiacrypt/xiang2016applying}. The MixColumn operation is treated as a single S-box. Finally, the model is solved using the Gurobi optimizer~\cite{gurobi}.
    \vspace{0.5em}

    \item[-] \textbf{Checking the Balance Property of the Linear Combination of Bits.} For any monomial $\pi_{\bm{u}_i}(\bm{s})$ in the ANF, its parity can be determined as follows:
    \[
    \sum_{i} \pi_{\bm{u}_i}(\bm{s}) = 
    \begin{cases}
    unknown & \text{if there exists } \bm{d} \in \mathbb{D}_{p} \text{ such that } \bm{u}_i \succeq \bm{d}, \\
    0 & \text{otherwise}.
    \end{cases}
    \]
    If all $\pi_{\bm{u}_i}(\bm{s})$ in the ANF is balanced, then the linear combination of bits $\mathcal{C}(\bm{b})$ is considered to be balanced.

\end{enumerate}
\begin{rrule}[\cite{dcc/LambinDF20}]\label{reducing-rule-1}
Given a monomial set $\mathbb{U}$, if there exist $\bm{u}_i, \bm{u}_j \in \mathbb{U}$ such that $\bm{u}_i \succeq \bm{u}_j$, then for all $\bm{u}_s$ such that $\bm{u}_j \succeq \bm{u}_s$, it always holds that $\bm{u}_i \succeq \bm{u}_s$. 
Therefore, it suffices to check the reduced set:
\[
\mathbb{U}' = \left\{ \bm{u}_i \in \mathbb{U} \ \middle|\ \nexists\ \bm{u}_j \in \mathbb{U} \text{ such that } \bm{u}_j \succeq \bm{u}_i \right\}.
\]
\end{rrule}
\par
The algorithm for checking the balance property of a linear combination of bits is presented in Algorithm~\ref{algorithm:check-balanced}.
\par
\begin{algorithm}
\caption{Search Key-Independnet Integral Distinguisher Based on Combination of Bits}
\label{algorithm:check-balanced}
\begin{algorithmic}[1]
\renewcommand{\algorithmicrequire}{\textbf{Input:}}
\renewcommand{\algorithmicensure}{\textbf{Output:}}
\Require A combination of ciphertext bits $\mathcal{C}(\bm{b})$, number of rounds $p$ and $q$, input division property $\bm{d}_0$
\Ensure \texttt{True} if $\mathcal{C}(\bm{b})$ is balanced; \texttt{False} otherwise
\State $\mathbb{U} \gets$ \textbf{BackwardExtension}$(\mathcal{C}, q)$
\State $\mathbb{U}' \gets$ \textbf{ApplyReducingRule}~\ref{reducing-rule-1}$(\mathbb{U})$
\For{each $\bm{u}_i \in \mathbb{U}'$}
    \State $\mathcal{M} \gets$ \textbf{Model}$(p, \bm{d}_0, \bm{u}_i)$
    \State $\mathcal{M}.\texttt{optimize}()$
    \If{$\mathcal{M}.\texttt{status}$ is \texttt{Optimal}}
        \State \Return \texttt{False}
    \EndIf
\EndFor
\State \Return \texttt{True}
\end{algorithmic}
\end{algorithm}

\par
\vspace{0.5em}
\noindent \textbf{Using Algorithm~\ref{algorithm:check-balanced} to Reproduce Distinguishers~\ref{dis:integral-distinguisher-7-round} and Distinguisher~\ref{dis:integral-distinguisher-8-round}} 
For Distinguisher~\ref{dis:integral-distinguisher-7-round}, we apply 1 round of backward monomial extension and 6 rounds of forward modeling. As shown in Equation~(\ref{equ:7-round-anf}), we verify whether each monomial in the ANF is balanced:
\begin{equation}
\label{equ:7-round-anf}
b_{4} \oplus b_{52} = s_{56}s_{57} \oplus s_{56} \oplus s_{57} \oplus s_{59} \oplus 1
\end{equation}

\noindent For Distinguisher~\ref{dis:integral-distinguisher-8-round}, we apply 2 rounds of backward monomial extension and 6 rounds of forward modeling. As shown in Equation~(\ref{equ:8-round-anf}), we similarly verify the balance property of each monomial.

\begin{equation}
\label{equ:8-round-anf}
\begin{aligned}
    &b_{28}\oplus b_{44}\oplus b_{60} \\
    = &s_{9}s_{10}k_{28} \oplus s_{8}s_{9}k_{29} \oplus s_{8}k_{28} \oplus s_{9}k_{28} \oplus s_{10}k_{28} \oplus  s_{8}k_{29} \oplus s_{9}k_{29} \oplus  s_{11}k_{29} \ \oplus \\
    &  k_{28}k_{29} \oplus s_{10} \oplus k_{31} \oplus k_{36}
\end{aligned}
\end{equation}
\noindent Finally, using the improved automated search model, we successfully identified Distinguisher~\ref{dis:integral-distinguisher-7-round} and~\ref{dis:integral-distinguisher-8-round}, which are consistent with the results obtained from the integral-neural distinguisher.
\par
\vspace{0.5em}
\noindent \textbf{Using Algorithm~\ref{algorithm:check-balanced} to Search for Longer Distinguishers} 
For \skinnyver{64}{64}, there are $2^{64}$ possible linear combinations of bits, making exhaustive search infeasible. However, based on the observed structure of the linear combinations in Distinguisher~\ref{dis:integral-distinguisher-7-round} and~\ref{dis:integral-distinguisher-8-round}, we find that each bit involved in the linear combination with balance property is always located at the same position within its respective column. Therefore, it suffices to examine only $4 \times (2^4 - 1)$ linear combinations of bits.
Using this strategy, we successfully obtain an integral distinguisher for 11-round \skinnyver{64}{64} and \skinnyver{128}{128} by activating only 60 plaintext bits, which refer to this as \textbf{Distinguisher~\ref{dis:integral-distinguisher-11-round}} and \textbf{Distinguisher~\ref{dis:integral-distinguisher-11-round-128}}. 
It is worth noting that to obtain an 11-round integral distinguisher for \skinnyver{64}{64}, Derbez \etal{} required 63 activated plaintext bits~\cite{fse/derbez2020increasing}.
\par
\begin{distinguisher}\label{dis:integral-distinguisher-11-round}
\emph{\textbf{(Key-Independent Integral Distinguisher Based on Linear Combination of Bits Against 11-round \skinnyver{64}{64}.)}} 
If the last 60 plaintext bits are activated, then the linear combinations of bits $b_i \oplus b_{i+16} \oplus b_{i+32}$ for $i \in \{16, 17, \dots, 31\}$ and $b_{12} \oplus b_{60}$ are balanced.
\end{distinguisher}

\begin{distinguisher}\label{dis:integral-distinguisher-11-round-128}
\emph{\textbf{(Key-Independent Integral Distinguisher Based on Linear Combination of Bits Against 11-round \skinnyver{128}{128}.)}} 
If the last 60 plaintext bits are activated, then the linear combinations of bits $b_i \oplus b_{i+16} \oplus b_{i+32}$ for $i \in \{26, \dots, 63\}$ and $b_{26} \oplus b_{122}$ are balanced.
\end{distinguisher}

\section{Key-Dependent Integral Distinguisher}
\label{sec:key-dependent-integral-distinguisher}
While our previous analysis has primarily focused on key-independent (i.e., deterministic) integral properties, probabilistic integral properties are also of significant value. In fact, such distinguishers correspond to cube testers, as introduced in~\cite{fse/AumassonDMS09}.
In this section, we demonstrate that neural networks are capable of learning probabilistic integral properties. Moreover, by employing the improved automated search model, we identify a higher-round key-dependent integral distinguisher for \skinnyver{64}{64}.


\subsection{Key-Dependent Integral Distinguisher Based on linear Combination of Bits}
\label{sub:key-depent-linear-combination}
In Section~\ref{sub:linear-combiniation-bit-distinguisher}, a 7-round integral-neural distinguisher using $\mathbb{VDS}$ with an accuracy close to 100\% is obtained by activating a single plaintext cell. In general, the accuracy of a neural distinguisher tends to decrease as the number of rounds increases.
This raises a natural question: What is the maximum number of rounds that an integral-neural distinguisher can achieve when only one plaintext cell is activated?
\par
\begin{figure}[htb]
    \centering
    \includegraphics[width=0.6\textwidth]{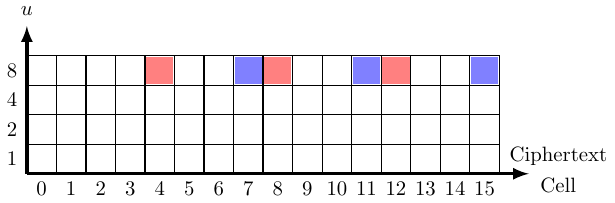}
    \caption{Bit sensitivity test of the 8-round integral-neural distinguisher for \skinnyver{64}{64} with the 15th plaintext cell activated.}
    \label{fig:bit-sensitivity-8-round-skinny64}
\end{figure}

\vspace{0.5em}
\noindent \textbf{The 8-Round Integral-Neural Distinguisher with One Activated Plaintext Cell}
The size of the $\mathbb{VDS}$ is reduced to 64, since $u \in \{\texttt{0b0001}, \texttt{0b0010},\allowbreak \texttt{0b0100}, \texttt{0b1000}\}$ and there are 16 cells in total.
An 8-round integral-neural distinguisher for \skinnyver{64}{64} is obtained by activating the 15th plaintext cell, as shown in Table~\ref{tab:8-round-64-distinguisher-cell-15}. A bit sensitivity test was then performed to analyze the learned features, and the results are presented in Fig.~\ref{fig:bit-sensitivity-8-round-skinny64}.

\begin{table}[htb]
\centering
\caption{Performance of the 8-round integral-neural distinguisher for \texttt{SKINNY-64-64} when the 15th plaintext cell is activated}
\label{tab:8-round-64-distinguisher-cell-15}
\begin{tblr}
{
  colspec = {Q[m,c] Q[m,c] Q[m,c] Q[m,c]},
  hline{1,Z} = {.08em},
  hline{2} = {.05em},
  cell{2}{3} = {gray9},
}
Activated Plaintext Cell & Acc & TPR   & TNR   \\
15th                     & 57.6\%   & 49.2\% & 65.9\% \\
\end{tblr}
\end{table}
\par

From Fig.~\ref{fig:bit-sensitivity-8-round-skinny64}, it can be observed that the 8-round integral-neural distinguisher relies on only 6 bits (highlighted in \fillcolor{blue!50} and \fillcolor{red!50}) from the $\mathbb{VDS}$. These bits correspond to the case where \( u = 8 \) in ciphertext cells \( \{4, 7, 8, 11, 12, 15\} \).
The 6 bits can be grouped into two categories: \fillcolor{red!50} and \fillcolor{blue!50}.
\begin{itemize}
    \item When the \fillcolor{red!50} bits are masked, the accuracy of the distinguisher decreases by 2.6\%.
    \item When the \fillcolor{blue!50} bits are masked, the accuracy of the distinguisher decreases by 4.7\%.
\end{itemize}

\par
\noindent \textbf{Retraining the Integral-Neural Distinguisher}
Given the abnormal TPR observed for the distinguisher in Table~\ref{tab:8-round-64-distinguisher-cell-15}, we retrain the integral-neural distinguisher using only the $\mathbb{VDS}$ corresponding to $u = 8$ in the ciphertext cells \( \{4, 8, 12\} \) and \( \{7, 11, 15\} \). The retraining results are presented in Table~\ref{tab:8-round-64-distinguisher-cell-15-u=8}.

\begin{table}[htb]
\centering
\caption{Integral-neural distinguisher for 8-round \texttt{SKINNY-64-64} using $\mathbb{VDS}$ entries corresponding to $u=8$ in ciphertext cells $\{4, 8, 12\}$ and $\{7, 11, 15\}$}
\label{tab:8-round-64-distinguisher-cell-15-u=8}
\SetTblrInner{colsep=6pt}
\begin{tblr}
{
  colspec = {Q[m,c] Q[m,c] Q[m,c] Q[m,c] Q[m,c] Q[m,c]},
  cell{2}{1,2} = {r=2}{},
  hline{1,Z} = {.08em},
  hline{2} = {.05em},
}
{Activated \\ Plaintext Cell} & $u$ & {Observed \\ Ciphertext Cells} & Acc & TPR & TNR \\
15th & 8 & $\{4, 8, 12\}$ & 56.3\% & 62.7\% & 49.9\% \\
15th & 8 & $\{7, 11, 15\}$ & 56.3\% & 62.7\% & 49.9\% \\
\end{tblr}
\end{table}

\par
We utilized $10^7$ positive instances to calculate the frequency of the equation \( b_{16} \oplus b_{32} \oplus b_{48} = 0 \), which corresponds to the case of $u = 8$ in ciphertext cells $\{4, 8, 12\}$. The analysis revealed that approximately 62.7\% of the instances satisfied this condition.
This frequency aligns precisely with the TPR of the integral-neural distinguisher reported in Table~\ref{tab:8-round-64-distinguisher-cell-15-u=8}, indicating that the neural network has successfully learned a probabilistic (i.e., key-dependent) balanced property. Based on this observation, we give a key-dependent integral distinguisher, as presented in Distinguisher~\ref{dis:key-dependent-integral-distinguisher-8-round}. A similar 8-round integral distinguisher also exists for \texttt{SKINNY-128-128}, as shown in Distinguisher~\ref{dis:key-dependent-integral-distinguisher-8-round-128}.

\begin{distinguisher}\label{dis:key-dependent-integral-distinguisher-8-round}
\emph{\textbf{(Key-Dependent Integral Distinguisher Based on Linear Combination of Bits Against 8-round \skinnyver{64}{64}.)}}
If the 15th plaintext cell is activated, then the linear combinations of bits $b_{16} \oplus b_{32} \oplus b_{48}$ and $b_{28} \oplus b_{44} \oplus b_{60}$ are probability balanced.
\end{distinguisher}

\begin{distinguisher}\label{dis:key-dependent-integral-distinguisher-8-round-128}
\emph{\textbf{(Key-Dependent Integral Distinguisher Based on Linear Combination of Bits Against 8-round \skinnyver{128}{128}.)}}
If the 15th plaintext cell is activated, then the linear combinations of bits $b_{33} \oplus b_{65} \oplus b_{97}$ and $b_{57} \oplus b_{89} \oplus b_{121}$ are probability balanced.
\end{distinguisher}

\subsection{Improved Automated Search Model for Key-Dependent Integral Distinguisher}
\label{sub:key-dependent-linear-combination}
To study the probability of the key-dependent integral distinguisher, we get ANF of the expression $b_{16} \oplus b_{32} \oplus b_{48}$ using a 2-round backward monomial extension. The resulting superpoly is shown in Equation~(\ref{equ:8-round-key-independent}), where the variable $\bm{s}$ denotes the input state of the last two rounds.

\noindent
\begin{equation}
\label{equ:8-round-key-independent}
\begin{aligned}
 & b_{16}\oplus b_{32}\oplus b_{48}\\ =&\underset{balanced}{\underbrace{s_{14}(1\oplus k_{16})\oplus s_{13}(k_{16}\oplus k_{17})\oplus s_{12}(k_{16}\oplus k_{17})\oplus s_{15}k_{17}\oplus s_{16}k_{17}\oplus k_{19}\oplus k_{48}}} \\
 \oplus &\underset{unknown}{\underbrace{s_{13}s_{14}k_{16}\oplus s_{12}s_{13}k_{17}}}.
\end{aligned}
\end{equation}
\par
Assuming the key is uniformly random, the probability that both \( k_{16} \) and \( k_{17} \) are equal to 0 is 0.25. In this case, the monomials \( s_{49}s_{50}k_{16} \oplus s_{50}s_{51}k_{17} \) becomes balanced.
When \( k_{16} \) and \( k_{17} \) are not both 0, which occurs with probability 0.75, the probability that monomials \( s_{49}s_{50}k_{16} \oplus s_{50}s_{51}k_{17} \) are balanced is 0.5.
Therefore, the overall probability that \( b_{16} \oplus b_{32} \oplus b_{48} \) is balanced can be estimated as:
\[
0.25 \times 1 + 0.75 \times 0.5 = 0.625,
\]
which closely matches the TPR of the integral-neural distinguisher reported in Table~\ref{tab:8-round-64-distinguisher-cell-15-u=8}. Notice, the estimated value of 0.625 deviates slightly from the experimental result of 0.627. This discrepancy arises because the probability that a monomial is $unknown$ or $balanced$ is not necessarily exactly 0.5.
\par
In previous work, the focus has primarily been on identifying key-independent integral distinguishers.
When a \textit{unknown} monomial is key-independent, it means there is no key bit in the monomial. In this case, there is no opportunity to make the \textit{unknown} monomial balanced by setting key bits to 0. Therefore, the monomial is certainly \textit{unknown}. If a key-independent \textit{unknown} monomial exists in the ANF, the entire ANF will be considered \textit{unknown}.
However, when specific key values are considered, it is possible for an $unknown$ monomial with key variable to become balanced.
As a result, a probabilistic (key-dependent) integral distinguisher may be obtained—resembling the notion of a weak-key integral distinguisher.
\par
\vspace{0.5em}
\noindent \textbf{Automatic Search for Key-Dependent Integral Distinguisher}
The algorithm for searching key-dependent integral distinguishers is shown in Algorithm~\ref{algorithm:calculate-probability}. Building upon Algorithm~\ref{algorithm:check-balanced}, we examine each \emph{unknown} monomial individually, collect the key bits involved in the monomial, and estimate the probability that the linear combination of bits is balanced when both key bits and constant plaintext bits are randomly assigned.
It is important to note that when key bits are present in the \emph{unknown} monomial, Reducing Rule~\ref{reducing-rule-1} becomes insufficient. For example, consider two monomials \( \pi_{\bm{u}_i}(\bm{s})\pi_{\bm{v}_i}(\bm{k}) \) and \( \pi_{\bm{u}_j}(\bm{s})\pi_{\bm{v}_j}(\bm{k}) \) such that \( \bm{u}_i \succeq \bm{u}_j \), \( \bm{v}_i \neq 0 \), and \( \bm{v}_j = 0 \). In this case, the monomial \( \pi_{\bm{u}_j}(\bm{s})\pi_{\bm{v}_j}(\bm{k}) \) must not be ignored, as it is certainly \emph{unknown}.
In contrast, if \( \bm{u}_i = \bm{u}_j \) and \( \bm{v}_i \preceq \bm{v}_j \) holds for all relevant monomials, then it is safe to disregard the pair \( (\bm{u}_j, \bm{v}_j) \). We summarize this new insight as Reducing Rule~\ref{reducing-rule-2}.
\par
\begin{rrule}\label{reducing-rule-2}
Given a set of monomials $\mathbb{M} = \{(\bm{u}_i, \bm{v}_i)\}$, the reduced set is defined as:
\[
\mathbb{M}' = \left\{ (\bm{u}_i, \bm{v}_i) \in \mathbb{M} \ \middle| \ \nexists\, (\bm{u}_j, \bm{v}_j) \in \mathbb{M} \text{ such that } \bm{u}_j \succeq \bm{u}_i \text{ and } \bm{v}_j \preceq \bm{v}_i \right\}.
\]
\end{rrule}

\begin{algorithm}
\caption{Search Key-Dependnet Integral Distinguisher Based on Combination of Bits}
\label{algorithm:calculate-probability}
\begin{algorithmic}[1]
\renewcommand{\algorithmicrequire}{\textbf{Input:}}
\renewcommand{\algorithmicensure}{\textbf{Output:}}
\Require Combination of ciphertext bits $\mathcal{C}(\bm{b})$, number of rounds $p$, $q$, input division property $\bm{d}_0$
\Ensure Estimated probability that $\mathcal{C}(\bm{b})$ is balanced
\State $\mathbb{M} \gets$ \textbf{BackwardExtension}$(\mathcal{C}, q)$
\State $\mathbb{M}' \gets$ \textbf{ApplyReducingRule}~\ref{reducing-rule-2}$(\mathbb{M})$
\State $\mathbb{V} \gets \emptyset$ \Comment{Set of key bit indices in the \emph{unknown} monomials}
\For{each $(\bm{u}_i, \bm{v}_i) \in \mathbb{M}'$}
    \If{$\bm{u}_i = \bm{0}$}
        \State \textbf{continue}
    \EndIf
    \State $\mathcal{M} \gets$ \textbf{Model}$(p, \bm{d}_0, \bm{u}_i)$
    \State $\mathcal{M}.\texttt{optimize}()$
    \If{$\mathcal{M}.\texttt{status}$ is \texttt{Optimal}}
        \If{$\bm{0} \in \mathbb{V}$}
            \State \Return $0$
        \EndIf
        \State $\mathbb{V}.\texttt{add}(\{ j \mid \bm{v}_i[j] = 1 \})$ \Comment{Collect key bits involved}
    \EndIf
\EndFor
\State \Return $2^{-|\mathbb{V}|} + (1 - 2^{-|\mathbb{V}|}) \times 0.5$ \Comment{Key bits = 0 with prob $2^{-|\mathbb{V}|}$}
\end{algorithmic}
\end{algorithm}
Notice, when Algorithm~\ref{algorithm:calculate-probability} returns a probability of 1, the integral distinguisher is in fact deterministically valid.
Using Algorithm~\ref{algorithm:calculate-probability}, we successfully identify Distinguisher~\ref{dis:integral-distinguisher-12-round}, which represents the longest known integral distinguisher for \skinnyver{64}{64} to date.
\par
\begin{distinguisher}\label{dis:integral-distinguisher-12-round}
\emph{\textbf{(Key-Dependent Integral Distinguisher Based on Linear Combination of Bits Against 12-round \skinnyver{64}{64}.)}}
If the last 60 plaintext bits are activated, then the probability that the linear combinations of bits $b_{16} \oplus b_{32} \oplus b_{48}$, $b_{24} \oplus b_{40} \oplus b_{56}$, and $b_{28} \oplus b_{44} \oplus b_{60}$ are balanced is approximately 0.625.
\end{distinguisher}

\subsection{Key-Dependent Integral Distinguisher Based on Nonlinear Combination of Bits}
\label{sub:key-dependent-nonlinear-combination}
In the early stages of integral cryptanalysis, both the integral property and the division property were originally studied at the byte level.
In this section, we shift our focus back to the byte level by using the division sequence $\mathbb{DS}$ corresponding to a ciphertext cell as the input to train the integral-neural distinguisher. The experimental results are presented in Table~\ref{tab:7-round-64-distinguisher-divisioin-sequence}.
\par
\begin{table}[H]
\centering
\caption{The integral-neural distinguisher for 7-round \texttt{SKINNY-64\allowbreak-64} using $\mathbb{DS}$}
\label{tab:7-round-64-distinguisher-divisioin-sequence}
\begin{tblr}
{
colspec={Q[m,c]Q[m,c]Q[m,c]Q[m,c]Q[m,c]},
hline{1,Z}={.08em},
hline{2}={.05em},
}
 {Activated \\ Plaintext Cell}  & {Observed \\ Ciphertext Cells} & Acc   & TPR   & TNR \\
15th     & 6th    & 76.3\% & 94.4\% & 58.3\%   \\
\end{tblr}
\end{table}
\par
\vspace{0.5em}
\noindent \textbf{Analyzing the Working Mechanism}
To further analyze the internal decision process of the integral-neural distinguisher, we adopt the method described in Section~\ref{sub:transform-distinguisher-to-boolean-function} to directly convert the trained model into a Boolean function. The conjunctive normal form (CNF) representation of the Boolean function consists of 490 clauses in total. For brevity, only the shortest clauses are listed below:
\\
$
(\bm{b}_{6}^{1}\lor \bm{b}_{6}^{2}\lor \bm{b}_{6}^{8}\lor \neg \bm{b}_{6}^{9}) \And
(\bm{b}_{6}^{1}\lor \neg \bm{b}_{6}^{2}\lor \bm{b}_{6}^{8}\lor \bm{b}_{6}^{9}) \And
(\bm{b}_{6}^{1}\lor \bm{b}_{6}^{4}\lor \bm{b}_{6}^{8}\lor \neg \bm{b}_{6}^{12}) \And
(\neg \bm{b}_{6}^1\lor \bm{b}_{6}^{4}\lor \bm{b}_{6}^{8}\lor \bm{b}_{6}^{12}) \And
(\bm{b}_{6}^{2}\lor \bm{b}_{6}^{8}\lor \neg \bm{b}_{6}^{9}\lor \bm{b}_{6}^{12}) \And
(\neg \bm{b}_{6}^{2}\lor \bm{b}_{6}^{8}\lor \bm{b}_{6}^{9}\lor \bm{b}_{6}^{12})
$
\par
For positive examples (i.e., the $\mathbb{DS}$ corresponding to a single ciphertext cell), every clause in the CNF must be satisfied. However, due to the large number of clauses, it is infeasible to directly infer the specific decision rules learned by the integral-neural distinguisher from the CNF representation.
To address this, we focus on the variables involved in each clause and evaluate their significance. Specifically, we use Algorithm~\ref{algorithm:check-balanced} to verify the integral property of various combinations of bits, such as:
\[
\bm{b}_{6}^{1} \oplus \bm{b}_{6}^{2} \oplus \bm{b}_{6}^{8} \oplus \bm{b}_{6}^{9}, \quad
\bm{b}_{6}^{1} \oplus \bm{b}_{6}^{4} \oplus \bm{b}_{6}^{8} \oplus \bm{b}_{6}^{12}, \quad
\bm{b}_{6}^{2} \oplus \bm{b}_{6}^{8} \oplus \bm{b}_{6}^{9} \oplus \bm{b}_{6}^{12}.
\]
\par
To facilitate understanding, we convert $b_{cth}^u$ into its representation as a nonlinear combination of bits, i.e.
\[
b_{24} \oplus b_{26} \oplus b_{27} \oplus b_{24}b_{27}, \quad
b_{24} \oplus b_{25} \oplus b_{27} \oplus b_{24}b_{25}, \quad
b_{24} \oplus b_{26} \oplus b_{24}b_{27} \oplus b_{24}b_{25}.
\]
We first perform one round of backward monomial extension on these three nonlinear combinations of bits, followed by six rounds of forward modeling. Each resulting monomial is then checked for balanc property. The results are presented in Equations~(\ref{equ:nonlinear-1}), (\ref{equ:nonlinear-2}), and (\ref{equ:nonlinear-3}).

\par
\begin{equation}\label{equ:nonlinear-1}
    \begin{aligned}
        & b_{24}\oplus b_{26}\oplus b_{27}\oplus b_{24}b_{27} \\
= &\underbrace{1 \oplus k_{30} \oplus k_{28} \oplus s_{8} \oplus k_{28}k_{31} \oplus s_{11}k_{31} \oplus s_{9}k_{31} \oplus s_{8}k_{31} \oplus s_{10}k_{28} \oplus s_{9}k_{28} \oplus s_{8}k_{28}}_{balanced}  \\
\oplus & \underbrace{s_{10}s_{11} \oplus s_{9}s_{11} \oplus s_{8}s_{11} \oplus s_{8}s_{10} \oplus s_{8}s_{9}k_{31} \oplus s_{10}s_{11}k_{28} \oplus s_{9}s_{11}k_{28} \oplus s_{8}s_{11}k_{28}}_{unknown} \\
\oplus &  \underbrace{s_{8}s_{10}k_{28} \oplus s_{9}s_{10}s_{11} \oplus s_{8}s_{9}s_{10} \oplus s_{9}s_{10}s_{11}k_{28} \oplus s_{8}s_{9}s_{10}k_{28}}_{unknown}
    \end{aligned}
\end{equation}

\begin{equation}\label{equ:nonlinear-2}
    \begin{aligned}
       & b_{24}\oplus b_{26}\oplus b_{24}b_{27}\oplus b_{24}b_{25} \\
= &\underbrace{k_{31} \oplus k_{30} \oplus k_{29} \oplus s_{11} \oplus s_{8} \oplus k_{28}k_{31} \oplus s_{11}k_{31} \oplus s_{9}k_{31} \oplus s_{8}k_{31} \oplus k_{28}k_{29}}_{balanced} \\
\oplus & \underbrace{s_{11}k_{29} \oplus s_{9}k_{29} \oplus s_{8}k_{29}}_{balanced}
\oplus \underbrace{s_{10}s_{11} \oplus s_{9}s_{11} \oplus s_{8}s_{11} \oplus s_{9}s_{10} \oplus s_{8}s_{10} \oplus s_{8}s_{9} }_{unknown}  \\
\oplus &  \underbrace{s_{8}s_{9}k_{31} \oplus s_{8}s_{9}k_{29} \oplus s_{10}s_{11}k_{28} \oplus s_{9}s_{11}k_{28} \oplus s_{8}s_{11}k_{28} \oplus s_{9}s_{10}k_{28} }_{unknown} \\
\oplus &  \underbrace{\oplus s_{8}s_{10}k_{28} \oplus s_{9}s_{10}s_{11} \oplus s_{8}s_{9}s_{10} \oplus s_{9}s_{10}s_{11}k_{28} \oplus s_{8}s_{9}s_{10}k_{28}}_{unknown}
    \end{aligned}
\end{equation}

\begin{equation}\label{equ:nonlinear-3}
    \begin{aligned}
    & b_{24}\oplus b_{25}\oplus b_{27}\oplus b_{24}b_{25} \\
= &  \underbrace{1 \oplus k_{31} \oplus s_{10} \oplus k_{28}k_{29} \oplus s_{11}k_{29} \oplus s_{9}k_{29} \oplus s_{8}k_{29} \oplus s_{10}k_{28} \oplus s_{9}k_{28} \oplus s_{8}k_{28}}_{balanced} \\
\oplus &  \underbrace{s_{8}s_{9}k_{29} \oplus s_{9}s_{10}k_{28}}_{uknown}
    \end{aligned}
\end{equation}
\par
From these three equations, we observe that only Equation~(\ref{equ:nonlinear-3}) allows the nonlinear combination of bits to become balanced by controling the key values, while Equations~(\ref{equ:nonlinear-1}) and~(\ref{equ:nonlinear-2}) do not, as they contain $unknown$ monomials that do not involve key bits.
Specifically, when $k_{28}$ and $k_{29}$ in Equation~(\ref{equ:nonlinear-3}) are set to 0, the probability that the nonlinear combination of bits is balanced becomes approximately 0.625. Therefore, we conclude that the neural network has learned certain probabilistic integral features from nonlinear combinations of bits.

\par
We enumerated all nonlinear combinations of bits for the 6th ciphertext cell, resulting in a total of $2^{2^4}$ possibilities. Using a process similar to Algorithm~\ref{algorithm:calculate-probability}, we calculated the probability that each of the $2^{16}$ nonlinear combinations of bits is balanced. Therefore, a key-dependent integral distinguisher for the nonlinear
combination of bits is proposed, as shown in Distinguisher~\ref{dis:7-round-nonlinear-distinguisher}.
\begin{distinguisher}\label{dis:7-round-nonlinear-distinguisher}
\emph{\textbf{(Key-Dependent Integral Distinguisher Based on Nonlinear combination of bits Against 7-round \skinnyver{64}{64}.)}}
If the 15th plaintext cell is activated, the probability that the nonlinear combination of bits $b_{24}\oplus b_{26}\oplus b_{24}b_{27}$ (related to $k_{28},k_{31}$) and $b_{24}\oplus b_{25}\oplus b_{27}\oplus b_{24}b_{25}$ (related to $k_{28},k_{29}$) are balanced is 0.625.
\end{distinguisher}
\par
Additionally, we derive an 11-round key-dependent integral distinguisher based on a nonlinear combination of bits against \skinnyver{64}{64}, as shown in Distinguisher~\ref{dis:11-round nonlinear-distinguisher}.
\begin{distinguisher}\label{dis:11-round nonlinear-distinguisher}
\emph{\textbf{(Key-Dependent Integral Distinguisher Based on Nonlinear Combination of Bits Against 11-round \skinnyver{64}{64}.)}} If the last 60 plaintext bits are activated, then the probability that the following nonlinear combinations of bits are balanced is approximately 0.625:
\[
b_{24} \oplus b_{26} \oplus b_{24}b_{27}  (\text{related to } k_{20}, k_{23}),  \text{and }
b_{24} \oplus b_{25} \oplus b_{27} \oplus b_{24}b_{25} (\text{related to } k_{20}, k_{21}).
\]
\end{distinguisher}

\par
The integral properties in Distinguishers~\ref{dis:7-round-nonlinear-distinguisher} and~\ref{dis:11-round nonlinear-distinguisher} correspond to the integral property defined in Definition~\ref{def:generalized-integral-property} where the function $g$ is nonlinear. We restrict the search space to all possible combinations within a single ciphertext cell. Specifically, we verify the balance of nonlinear combinations of bits within a single ciphertext cell for \skinnyver{64}{64}, where the number of calls to the automated search model is \( 2^{2^4} \).
However, extending this approach to \skinnyver{128}{128} would require evaluating \( 2^{2^8} \) combinations, which makes automated search computationally infeasible at this scale.

\section{Improved Key Recovery Attack}
\label{sec:improved-attack}
At CRYPTO 2016, Beierle \etal{} introduced the \skinny{} family of tweakable block ciphers and evaluated its resistance to integral attacks~\cite{crypto/BeierleJKL0PSSS16}. They identified a 6-round integral distinguisher using single-cell plaintext activation. By extending this distinguisher 4 rounds backward and 4 rounds forward, the authors successfully mounted 14-round key recovery attacks against both \skinnyver{64}{64} and \skinnyver{128}{128}.
Subsequent work by Zhang \etal{}~\cite{tosc/ZhangCGP19} demonstrated a 16-round integral attack on \skinnyver{128}{128}, using a 7-round distinguisher extended by 3 backward and 6 forward rounds. However, their analysis focused solely on the key guessing space and did not fully evaluate the time complexity.
More recently, Hosein \etal{} proposed an 18-round attack by constructing an integral distinguisher derived from a zero-correlation distinguisher under a chosen-tweak model. Their attack targets the 60-bit and 120-bit key variants of \skinnyver{64}{64} and \skinnyver{128}{128}, respectively~\cite{tosc/HadipourGSE24}.

\subsection{Key Recovery Attack Using Key-Independent Integral Distinguisher}
\label{15-skinny64-key-recovery-single-key}
The longest integral distinguisher does not necessarily lead to the most effective key-recovery attack. We use a 7-round integral distinguisher based on a linear combination of bits to perform a key-recovery attack. The detailed procedure is described as follows.
\par
\begin{figure}[htb]
    \centering
    \includegraphics[width=0.5\textwidth]{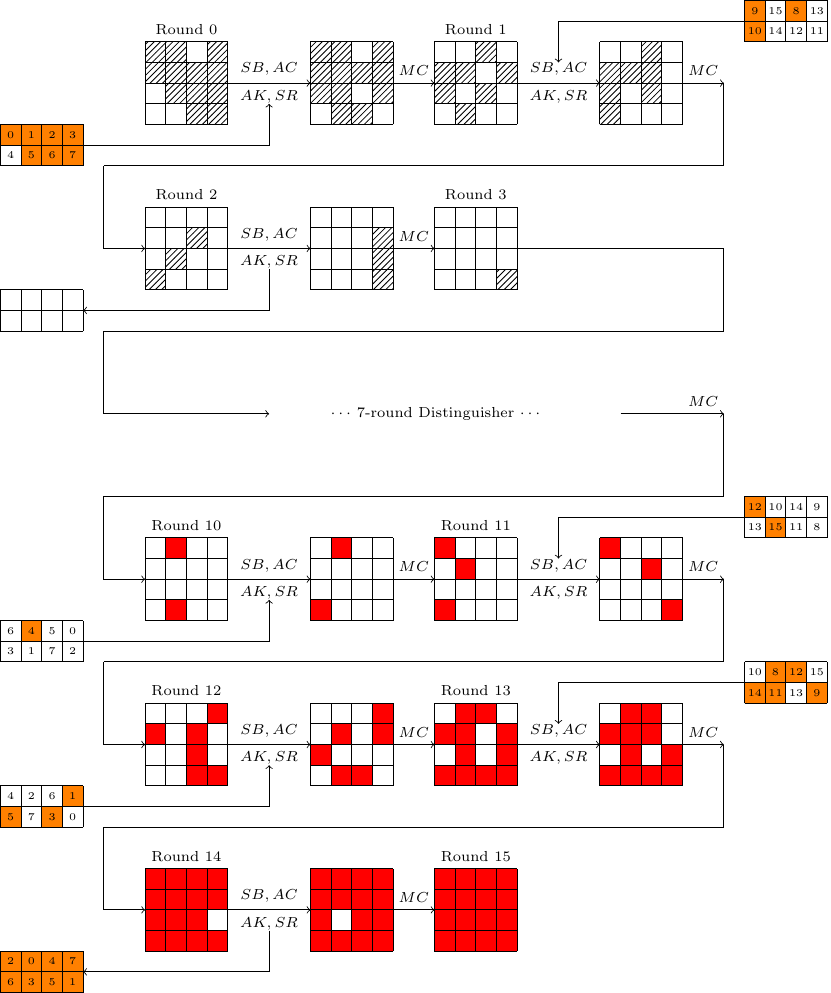}
    \caption{15-round key recovery using key-independent integral distinguisher for \skinnyver{n}{n} in single tweakey setting}
    \label{15-round-key-recovery-attack-single-key}
\end{figure}

\vspace{0.5em}
\noindent \textbf{The Key Recovery Attack for \skinnyver{n}{n}.}
The overall procedure of the key-recovery attack is illustrated in Figure~\ref{15-round-key-recovery-attack-single-key}
The target cipher $E$ is divided into three parts: $E_0$ covers rounds $(0 \rightarrow \ldots \rightarrow 3)$, $E_1$ (\textbf{Distinguisher~\ref{dis:integral-distinguisher-7-round}}) covers rounds $(3 \rightarrow \ldots \rightarrow 10)$, and $E_2$ covers rounds $(10 \rightarrow \ldots \rightarrow 15)$. The strategy proceeds as follows:
\begin{enumerate}
\item Obtain a pair of plaintext-ciphertext $(P, C)$.
\item Guess 15-cell involved keys,
\begin{itemize}
    \item Determine the value of $P$ in \Xcell of $E_0$. Encrypt the plaintext $P$ using the guessed key to obtain the ciphertext after the 3rd round. Then, traverse the 15th cell of the 3rd-round ciphertext to obtain a ciphertext set. Next, perform 3 rounds of decryption to recover the initial plaintext set. Finally, execute 15 rounds of encryption to obtain the final ciphertext set.

    \item Determine the values in \fillcolor{red} of round 10: partially decrypt each ciphertext using the guessed involved keys. Check whether the involved ciphertext bits are balanced and obtain the candidate round keys that pass the test.

\end{itemize}
\end{enumerate}
\par

\par
\noindent \textbf{Complexity Analysis}. In round 0, a total of 12 cells are activated, resulting in a data complexity of $2^{12c}$. A total of 16 plaintexts and ciphertexts are used, and the entire attack process involves keys from 15 cells and encryption/decryption of 59 cells. Therefore, the time complexity is calculated as $2^{c} \times 2^{15 \times c} \times \frac{59}{16 \times 15} = 2^{16c-2.024}$.
\par
\vspace{0.5em}
\noindent \textbf{The Key Recovery Attack for \skinnyver{n}{2n} and \skinnyver{n}{3n}}. Since the 7-round integral distinguisher is independent of the key, it can be applied to \texttt{SKINNY-n} with different key lengths.
\begin{enumerate}
    \item [-] Using a similar procedure, we can perform a 17-round key recovery attack on \skinnyver{n}{2n}. The attack process is shown in Appendix~\ref{app:key-attack-skinny-2n-3n} Fig.~\ref{17-round-key-recovery-attack}.The data complexity of $2^{12c}$ and the time complexity is  $2^{c} \times 2^{31 \times c} \times \frac{59+32}{16 \times 17} =  2^{32c-1.579}$.
    \vspace{0.5em}
    \item [-] Using a similar procedure, we can perform a 19-round key recovery attack on \skinnyver{n}{3n}. The attack process is shown in Appendix~\ref{app:key-attack-skinny-2n-3n} Fig.~\ref{19-round-key-recovery-attack}. The data complexity of $2^{12c}$ and the time complexity is $2^{c} \times 2^{47 \times c} \times \frac{59+64}{16 \times 19} =  2^{48c-1.305}$.
\end{enumerate}

\subsection{Key Recovery Attack Using Key-Dependent Integral Distinguisher}
\label{15-skinny64-key-recovery-weak-key}
In Section~\ref{15-skinny64-key-recovery-single-key}, the key recovery attack uses an integral distinguisher based on the linear combination of bits (\textbf{Distinguisher}~\ref{dis:integral-distinguisher-7-round}) involving two cells, which results in the need to guess two key bytes in the 11th round. When using a 8-round key-dependent integral distinguisher (\textbf{Distinguisher}~\ref{dis:key-dependent-integral-distinguisher-8-round}), fewer key bytes need to be guessed. Note that the attack is only effective when the two key bytes involved in the distinguisher are set to zero.
\par
\begin{figure}[htb]
    \centering
    \includegraphics[width=0.5\textwidth]{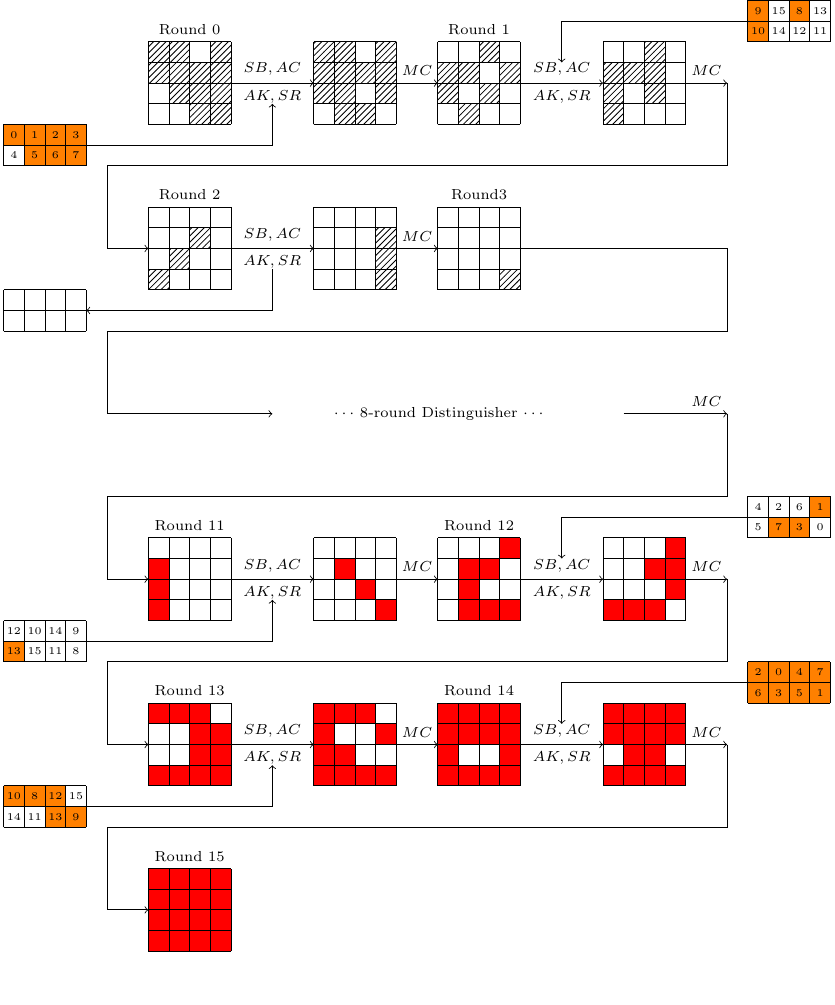}
    \caption{15-round key recovery using key-dependent integral distinguisher for \skinnyver{n}{n} in single tweakey setting}
    \label{15-round-key-recovery-attack-weak-key}
\end{figure}

\vspace{0.5em}
\noindent \textbf{The Complexity Analysis of Key Recovery Attack for \skinnyver{n}{n}}. The overall procedure of the key-recovery attack is illustrated in Figure~\ref{15-round-key-recovery-attack-weak-key}. In round 0, a total of 12 cells are activated, resulting in a data complexity of $2^{12c}$. A total of 16 plaintexts and ciphertexts are used, and the entire attack process involves keys from 15 cells and encryption/decryption of 57 cells. Therefore, the time complexity is calculated as $2^{c} \times 2^{13 \times c} \times \frac{57}{16 \times 15} = 2^{14c-2.074}$.
\par
\vspace{0.5em}
\noindent \textbf{The Key Recovery Attack for \skinnyver{n}{2n} and \skinnyver{n}{3n}}. For different parameter settings of the \texttt{SKINNY}, the 8-round key-dependent integral distinguishers involve different key positions, but the number of involved key bytes remains the same.
\begin{enumerate}
    \item [-] Using a similar procedure, we can perform a 17-round key recovery attack on \skinnyver{n}{2n}. The attack process is shown in Appendix~\ref{app:key-attack-skinny-2n-3n-weak} Fig.~\ref{17-round-key-recovery-attack-weak}. The data complexity of $2^{12c}$ and the time complexity is  $2^{c} \times 2^{29 \times c} \times \frac{57+32}{16 \times 17} =  2^{30c-1.612}$.
    \vspace{0.5em}
    \item [-] Using a similar  procedure, we can perform a 19-round key recovery attack on \skinnyver{n}{3n}. The attack process is shown in Appendix~\ref{app:key-attack-skinny-2n-3n-weak} Fig.~\ref{19-round-key-recovery-attack-weak}. The data complexity of $2^{12c}$ and the time complexity is $2^{c} \times 2^{45 \times c} \times \frac{57+64}{16 \times 19} =  2^{46c-1.329}$.
\end{enumerate}

\section{Conclusion}
\label{sec:conclusion}
This paper explores the application of neural networks in integral cryptanalysis and presents several significant advancements.
By comparing the results obtained through neural networks and classical methods, we observe that existing automated search models are inaccurate. Using an improved automated search model, we extend the length of the integral distinguishers for \texttt{SKINNY}. 
In addition to discovering longer integral distinguishers, we also improve key-recovery attacks. Specifically, the short-round distinguisher learned by the neural network enables a longer-round key-recovery attack on \texttt{SKINNY} through a combination of backward decryption and forward encryption.
These results demonstrate that artificial intelligence techniques can offer valuable support in advancing cryptanalysis. Unfortunately, identifying integral distinguishers requires an exhaustive enumeration of all possible mapping function. Exploring more efficient search strategies remains an important direction for future research.

\bibliographystyle{splncs04}
\bibliography{refbib}

\clearpage
\appendix
\appendix
\section{The Algorithm of Bit Sensitivity Test}
\begin{algorithm}
\caption{Bit Sensitivity Test~\cite{cj/ChenSY23}}\label{alg:bit_sensitively_test}
\begin{algorithmic}[1]
\renewcommand{\algorithmicrequire}{\textbf{Input:}}
\renewcommand{\algorithmicensure}{\textbf{Output:}}
\Require Test dataset $X=\{X_0,\cdots,X_{N-1}\}$, where half of the dataset is positive, and half is negative instances;
$n$: the size of an instance $X_{i}$; $\mathcal{ND}$: neural distinguisher;
\Ensure An array of the decrease in accuracy $\text{Acc}_{dec}$ for each bit;
\State $\text{Acc}_{orig} \gets \mathcal{ND}(X)$;
\State $\text{Acc}_{dec} \gets \{\}$;
\State $X_{masked} \gets \{\} $;
\For{$i = 0$ to $n-1$}
    \For{$j = 0$ to $N$}
        \State Generate a random mask $\eta \in \{0, 1\}$;
        \State $X_{masked} \gets X^j \oplus ( \eta \ll i )$; \Comment{Randomize bit $i$}
    \EndFor
    \State  Add $\text{Acc}_{orig} - \mathcal{ND}(X_{masked}) $ to $\text{Acc}_{dec}$;
\EndFor
\State \textbf{return} $\text{Acc}_{dec}$;
\end{algorithmic}
\end{algorithm}

\section{Improced Key Recovery Attack}

\subsection{Improved Key Recovery Attack Using Key-Independent Integral Distinguisher}
\label{app:key-attack-skinny-2n-3n}
\begin{figure}[htb]
    \centering
    \includegraphics[width=0.6\textwidth]{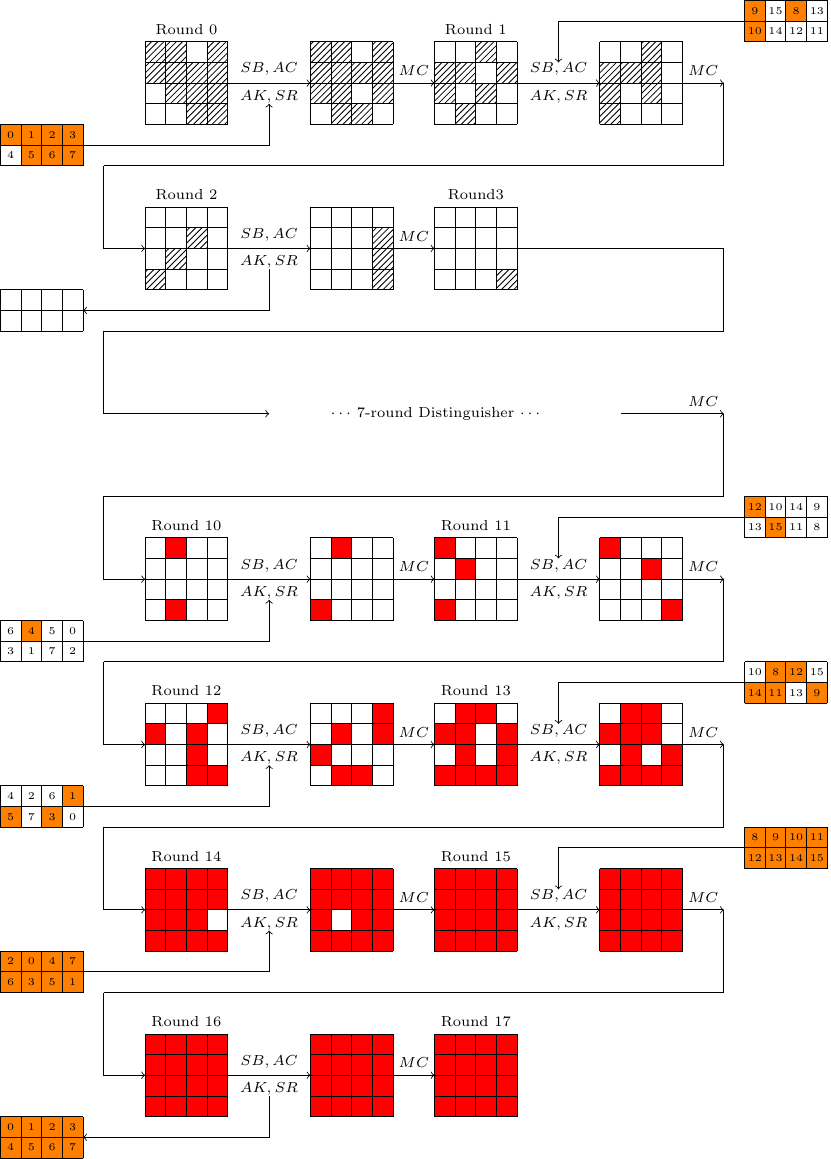}
    \caption{17-round key recovery using key-independent integral distinguisher for \skinnyver{n}{2n} in single tweakey setting}
    \label{17-round-key-recovery-attack}
\end{figure}

\begin{figure}[htb]
    \centering
    \includegraphics[width=0.6\textwidth]{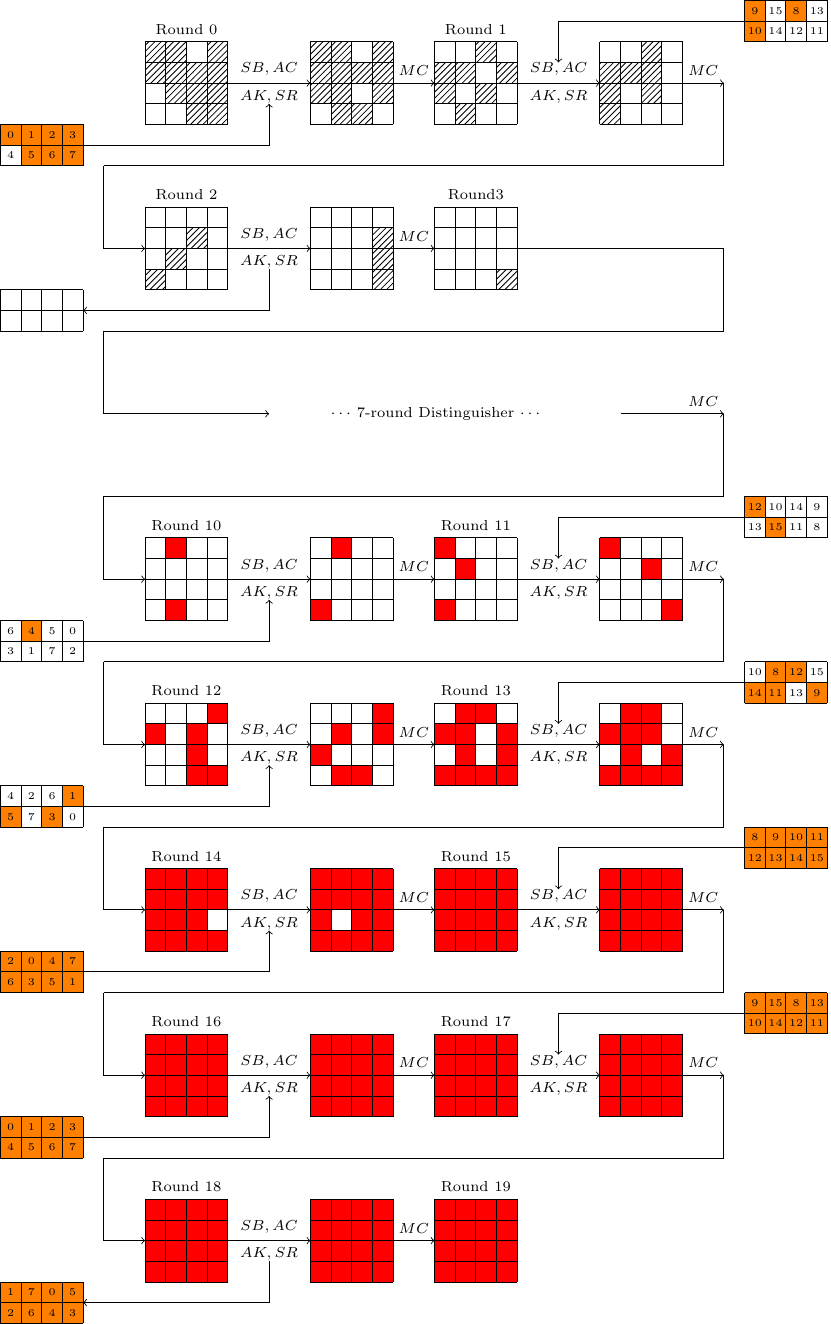}
    \caption{19-round key recovery using key-independent integral distinguisher for \skinnyver{n}{3n} in single tweakey setting}
    \label{19-round-key-recovery-attack}
\end{figure}

\clearpage

\subsection{Improved Key Recovery Attack Using Key-Dependent Integral Distinguisher}
\label{app:key-attack-skinny-2n-3n-weak}
\begin{figure}[htb]
    \centering
    \includegraphics[width=0.6\textwidth]{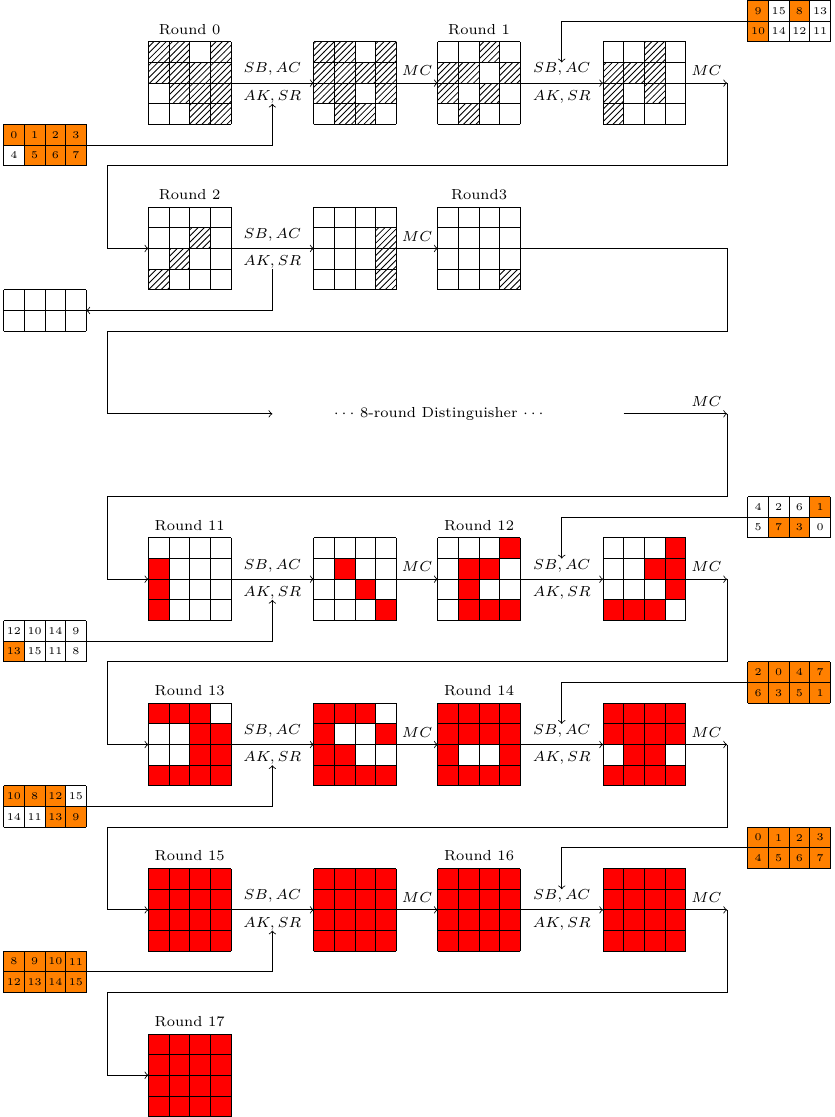}
    \caption{17-round key recovery using key-dependent integral distinguisher for \skinnyver{n}{2n} in single tweakey setting}
    \label{17-round-key-recovery-attack-weak}
\end{figure}

\begin{figure}[htb]
    \centering
    \includegraphics[width=0.6\textwidth]{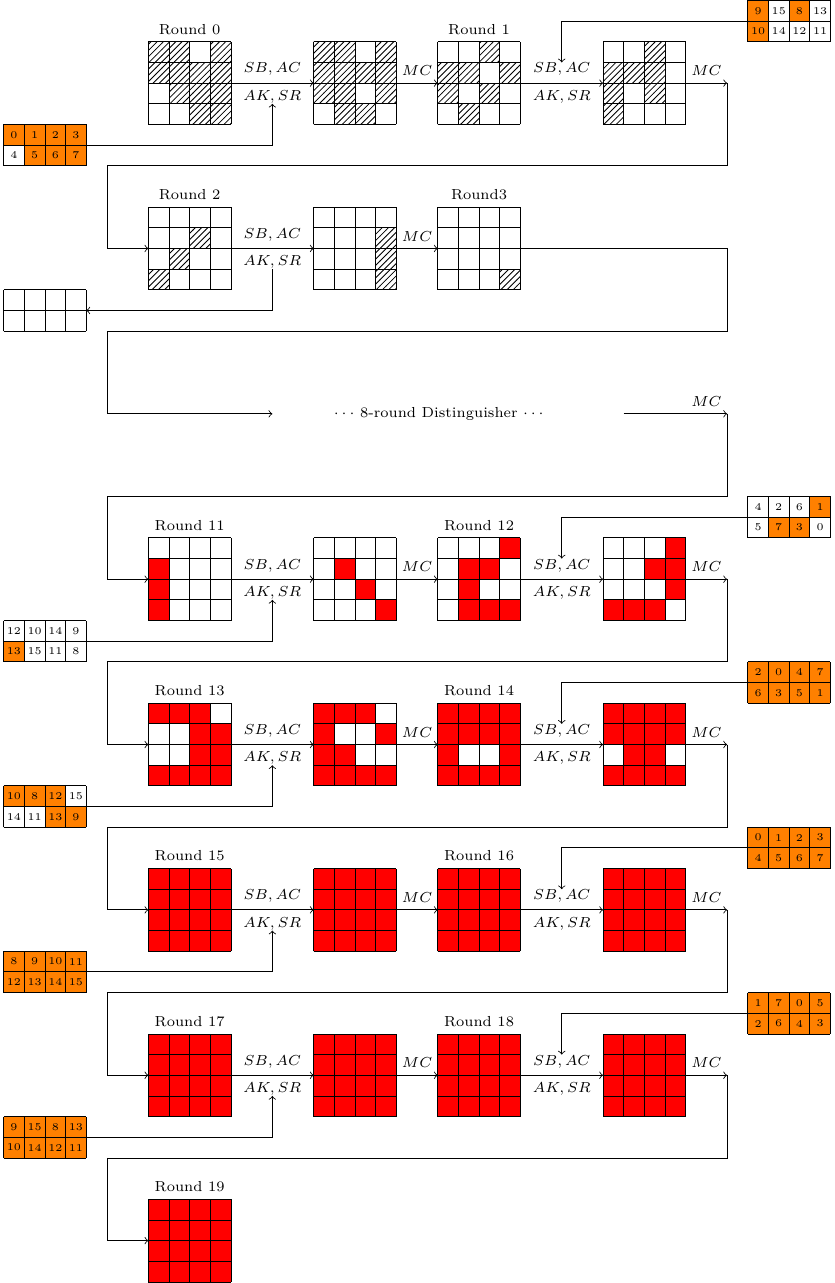}
    \caption{19-round key recovery using key-dependent integral distinguisher for \skinnyver{n}{3n} in single tweakey setting}
    \label{19-round-key-recovery-attack-weak}
\end{figure}



\end{document}